\documentclass[twocolumn]{aastex631}
\usepackage[utf8]{inputenc}
\usepackage{graphicx,psfrag}
\usepackage{mathrsfs}
\usepackage{amsmath,amsfonts,amssymb}
\usepackage{multirow}
\usepackage{diagbox}
\usepackage{comment}
\usepackage{xcolor}
\usepackage{enumerate}
\usepackage{booktabs}
\usepackage[normalem]{ulem}
\usepackage{hyperref}
\usepackage{mathtools}

\usepackage{siunitx}
\usepackage{float}
\usepackage{textgreek}
\hypersetup{
    colorlinks = true,
    linkcolor = {blue},
    citecolor = {blue},
    urlcolor = {blue},
    linkbordercolor = {white},
    }
\usepackage{color}
\usepackage{bigints}
\definecolor{cyan}{rgb}{0,0.9,0.9}
\definecolor{orange}{rgb}{0.9,0.5,0}
\definecolor{magenta}{rgb}{1,0,1}
\definecolor{purple}{rgb}{0.8,0.4,0.8}
\definecolor{gray}{rgb}{0.8242,0.8242,0.8242}
\definecolor{green}{rgb}{0.,0.8,0.}
\usepackage{subfigure}
\usepackage{tabularx}

\DeclareMathAlphabet\mathbfcal{OMS}{cmsy}{b}{n}

\usepackage[normalem]{ulem}

\begin{document}

\title{Consistent Treatment of Muons in Binary Neutron Star Mergers}

\author{Henrique \surname{Gieg}}
\affiliation{Institut f\"ur Physik und Astronomie, Universit\"at Potsdam, Haus 28, Karl-Liebknecht-Str. 24/25, 14476, Potsdam, Germany}
\author{Ramon \surname{Jaeger}}
\affiliation{Institut f\"ur Physik und Astronomie, Universit\"at Potsdam, Haus 28, Karl-Liebknecht-Str. 24/25, 14476, Potsdam, Germany}
\author{Maximiliano \surname{Ujevic}}
\affiliation{Centro de Ci\^encias Naturais e Humanas, Universidade Federal do ABC, 09210-170, Santo Andr\'e, S\~ao Paulo, Brazil}
\author{Tim \surname{Dietrich}}
\affiliation{Institut f\"ur Physik und Astronomie, Universit\"at Potsdam, Haus 28, Karl-Liebknecht-Str. 24/25, 14476, Potsdam, Germany}
\affiliation{Max Planck Institute for Gravitational Physics (Albert Einstein Institute), Am M\"uhlenberg 1, Potsdam 14476, Germany}

\begin{abstract}
We present a set of numerical-relativity binary neutron star merger simulations incorporating muons and muonic reactions for two baseline baryonic equations-of-state. In order to investigate the possible impact of muons and muonic weak reactions, we treat neutrinos with a gray (energy-independent) truncated moments scheme and an implicit-explicit time integrator. Newly computed neutrino rates are employed within the full kinematics approach for a set of relevant reactions, and pair-processes are modeled via opacities computed using reaction kernels, that allow a consistent treatment of neutrino interaction rates. We find that equilibration between matter and radiation is successfully captured by a novel two timescales approach. Of astrophysical interest is the general agreement between our muonic and non-muonic results regarding the remnant evolution, disk and outflow properties. Average electron fractions, asymptotic velocities and temperatures are different by less than $\sim 6\%$, while the main impact of muons is a reduction in ejecta masses by at most $\sim 17\%$. Therefore, based on our findings, accounting for the presence of muons and muonic reactions might result much less severe consequences regarding nucleosynthetic yields and electromagnetic counterparts than previously reported in the literature.
\end{abstract}
\date{\today}

\section{Introduction}
 Thermodynamical properties of ejected material in binary neutron star (BNS) mergers play a crucial role in the interpretation of electromagnetic counterparts of such events, e.g., the kilonova AT2017gfo (\cite{LIGOScientific:2017ync, Arcavi:2017xiz, Coulter:2017wya, Lipunov:2017dwd, Tanvir:2017pws, Valenti:2017ngx, Smartt:2017fuw}), follow-up of the gravitational-wave (GW) event GW170817 (\cite{TheLIGOScientific:2017qsa}). The current understanding is that kilonovae are powered by radioactive decay of heavy elements formed by r-process nucleosynthesis in the ejecta, e.g.,~(\cite{Rosswog:2017sdn, Kasen:2017sxr, Metzger:2019zeh, Arcones:2022jer}), while the multiple components of a kilonova signal are strongly correlated to the electron fraction $Y_e$ of the outflows (\cite{Drout:2017ijr, Evans:2017mmy, Nicholl:2017ahq, Perego:2017wtu}). In neutron-rich environments $Y_e \lesssim 0.25$, lanthanides are abundant and the medium is highly opaque (\cite{Kasen:2013xka, Barnes:2013wka, Tanaka:2013ana}), leading to a ``red'' kilonova component. Conversely, in neutron-poor media $Y_e \gtrsim 0.25$, lanthanides are less abundant and the ``blue'' component emerges.

From the theoretical point, assessment of the ejecta properties is only possible via numerical-relativity simulations of merging BNSs including relevant physics, such as general relativistic hydrodynamics (\cite{Shibata:2003ga, Rezzolla:2010fd, Bauswein:2013yna, Hotokezaka:2013iia, Dietrich:2015iva, Dietrich:2015pxa, Bernuzzi:2016pie, Radice:2013hxh, Radice:2013xpa, Radice:2017zta}, general relativistic ideal magnetohydrodynamics (\cite{Giacomazzo:2007ti, Mosta:2013gwu, Siegel:2014aaa,Kiuchi:2015sga,Ciolfi:2020cpf,Palenzuela:2021gdo, Kiuchi:2023obe, Neuweiler:2024jae, Gutierrez:2025gkx}), and neutrino transport
(\cite{Shibata:2011kx,Foucart:2016rxm, Radice:2021jtw, Schianchi:2023uky, Musolino:2023pao, Neuweiler:2025klw, Daszuta:2026szb}), the latter being vital for the compositional and thermal dynamics of matter. However, most of the existing BNS simulations rely on the assumption that matter contains only electrons ($e^-$) and positrons ($e^+$) as charged leptons. Consequently, neutrino transport is restricted to 3 neutrino species (hereafter, 3-$\nu$), namely, electron (anti)neutrinos $\nu_e~(\bar\nu_e)$ and heavy lepton neutrinos $\nu_x$, collectively representing muon (anti)neutrinos $\nu_\mu~(\bar\nu_\mu)$ and taon (anti)neutrinos $\nu_\tau~(\bar\nu_\tau)$. Such an approximation is, nonetheless, inaccurate, as in dense media the chemical potential of electrons $\mu_e$ may exceed the rest-mass of muons $m_\mu \approx 105.7~{\rm MeV}$, marking the onset of (anti)muons $\mu^-$ ($\mu^+$). In this case, the thermodynamical state of matter is determined by the temperature $T$, baryon number density $n_b$, net electron fraction $Y_e$ and, additionally, by the net muon fraction $Y_\mu$. Most importantly, in this scenario, neutrinos transport must address the evolution of (anti)muon neutrinos, demanding at least a 5 species (5-$\nu$) treatment, where, due to the substantially higher mass of taons, taon (anti)neutrinos may still be treated as a single representative species $\nu_x$.

Much more recently, the possible astrophysical implications of a consistent treatment of charged leptons in matter and its neutrino counterparts has been considered, first in the context of Core-Collapse Supernovae simulations, showing that the production of muons is significant for the dynamics of the collapse, via softening of the equation of state (EOS) and for the neutrino heating efficiency (\cite{Bollig:2017lki, Fischer:2020vie}). In BNS simulations, \cite{Loffredo:2022prq} evaluated the impact of muons in the post-merger remnant via post-processing of 3-$\nu$ simulation data. Then, in \cite{Gieg:2024jxs}, the first BNS merger simulations with muons and muonic weak interactions, albeit with a simplified transport (neutrinos leakage scheme) and neutrino rates (within the elastic approximation), showing that de-muonization of low density, cold material is a natural consequence of muonic $\beta$-reactions. Furthermore, ejecta masses from 3-$\nu$ simulations may be overestimated by a factor of 2, and that the softening of the EOS induced by muons lead to enhanced stabilization of the post-merger remnant by rotational support. Still, due to limitations of the approximate transport scheme, no strong imprints in the composition and entropy of the ejecta could be verified. Finally, substantial advances were presented in \cite{Ng:2024zve}, where 3- and 5-$\nu$ simulations were performed with a truncated moments transport scheme (M1 scheme) for neutrinos and a more realistic set of neutrinos rates. There, the authors show a similar factor of 2 smaller ejecta mass for the muonic simulations and, most notably, that muonic reactions noticeably cool the remnant, leading to dramatically smaller $Y_e$ in the ejecta over the simulation time, with consequences for the nucleosynthetic yields. However, some technical improvements for the treatment of pair-processes are necessary, given their fundamental importance for the dynamics of muonic leptons.

In this \textit{Letter} we present a set of numerical-relativity simulations of BNS mergers in the 3- and 5-$\nu$ case with a gray M1 scheme and realistic neutrino rates, including pair-processes via reaction kernels, allowing the construction of finite and well-behaved energy-integrated opacities, and, for the first time in the BNS context, the inclusion of (inverse) muon decay.

\section{Methods and Setups}
We present a set of four numerical-relativity BNS merger simulations using the \texttt{BAM} code (\cite{Bruegmann:2006ulg, Thierfelder:2011yi}, including a gray (energy-integrated) M1 moment scheme for neutrinos (see, e.g., \cite{Schianchi:2023uky} for details of our implementation), extended to account for 5$-\nu$ species. The time evolution of stiff radiation fields and the coupling with matter is performed via an implicit-explicit (IMEX) scheme (\cite{Izquierdo:2022eaz}), using the explicit RK(4,4) (4 stages, 4th order accurate) and the associated implicit I42L (4 stages, 2nd order accurate, L-stable) method. Implementation details for the IMEX scheme are left for an upcoming article (\cite{Jaeger:2026abc}).

The geometry is evolved explicitly via the Z4c formalism (\cite{Hilditch:2012fp}), while matter is evolved explicitly as an ideal fluid, whose conserved variables are governed by the general-relativistic hydrodynamics (GRHD) equations in the Valencia formulation (\cite{Font:2008fka}). For the 5-$\nu$ simulations, we evolve separately the net electron fraction $Y_e$ and the net muon fraction $Y_\mu$ according to
\begin{eqnarray}\label{eq:lep-bal}
    \partial_0(\sqrt{\gamma}W\rho Y_{l}) + \partial_i(\sqrt{\gamma} W \rho Y_l u^i) = m_b \mathcal{S}_{Y_l},
\end{eqnarray}
$l \in \{e, \mu\}$, $\gamma$ is the determinant of the spatial metric, $W$ is the Lorentz factor of a fluid element, $u^i$, $i = 1,2,3$, is the spatial component of the 4-velocity, $\rho = m_b n_b$ the rest-mass density, $m_b$ the baryon mass constant, $n_b$ the baryon number density, and $\mathcal{S}_{Y_l}$ is the number source-term for the lepton $l$, given by the difference of net emissions of antineutrinos and neutrinos of flavor $l$.

The GRHD system is closed by an EOS, whose construction is described in \cite{Gieg:2024jxs}. In this work we choose to parameterize the EOS and neutrino rates in terms of the rest-mass density $\rho$, temperature $T$, net electron fraction $Y_e \in [0.01, 0.5]$ with $50$ points and linear uniform stride, and the net muon fraction $Y_\mu \in [1\times10^{-4}, 0.1]$ with $61$ points and $\log_{10}$ uniform stride. Whenever necessary, the proton fraction $Y_p$ parameterizing the baryonic sector is obtained by enforcing local charge neutrality, i.e., $Y_p = Y_e + Y_\mu$. For the high-resolution shock-capturing scheme employed in the hydrodynamics code, we adopt the WENOZ reconstruction~(\cite{Borges:2008}) and the HLLE Riemann solver (\cite{Harten:1983, Einfeldt:1988}).

Our four simulated setups are equal-mass systems with total mass $M = 2.5~M_\odot$ modeled by the SFHo~(\cite{Steiner:2012rk}) and DD2~(\cite{Hempel:2009mc}) baryonic baselines ``dressed'' with photons, electrons and positrons for the 3-$\nu$ case, and additionally with muons and antimuons for 5-$\nu$ case. The numerical domain consists of 8 nested Cartesian boxes (levels), with a 2:1 spacing refinement from coarser to finer. Moving levels track the motion of the stars, and the grid spacing in the finest level is $\Delta x \approx 185~{\rm m}$. Our CFL (Courant-Friedrichs-Lewy) number is set to $0.25$. To focus on the neutrinos physics, we neglect magnetic fields. The neutrino processes employed in this work are listed in Table~\ref{tab:reactions}. Absorption opacities (corrected for stimulated absorption) of the $\beta$-reactions (a) - (f) follow the full kinematics approach of~\cite{Guo:2020tgx}, iso-energetic scattering opacities on free nucleons (g) and nuclear clusters (h) receive all relevant energy-dependent corrections~(\cite{Horowitz:1997neutrino, Mezzacappa:1993gm, Burrows:2006neutrino, Ardevol-Pulpillo:2018btx}). The isotropic annihilation kernels for pair-processes opacities of electron-positron pair annihilation (i) (\cite{Bruenn:1985stellar, Pons:1998st}), nucleon-nucleon bremsstrahlung (j) (\cite{Hannestad:1997gc}) and inverse (anti)muon decay (k) and (l) (\cite{Guo:2020tgx}) are used to tabulate absorption opacities with the approximation that pairing neutrinos are in local thermodynamical equilibrium (LTE) with matter. A detailed description of our methods to obtain well-behaved gray opacities and the associated emission rates are found in Sec.~\ref{sec:nu-rates} of the Supplementary Material. Finally, we extend the neutrino equilibration prescription of~\cite{Perego:2019adq} to the $5$-$\nu$ case, henceforth called the two timescales approach, which is also described in Sec.~\ref{sec:equili} of the Supplementary Material.

\begin{table}[t]
    \caption{Neutrino reactions considered in this work. Note that pair processes (i) and (j) are only considered for $\nu_x,~\nu_\mu,~\bar\nu_\mu$. $N$ is a free nucleon, while $A$ represents a heavy nuclear cluster or an alpha particle. All neutrino species participate on elastic scatterings (g) and (h).}
\begin{tabularx}{\columnwidth}{ll}
\hline
(a) $\nu_e + n \leftrightarrow p + e^-$                &  \hspace{0.8cm} (g) $\nu + N \rightarrow \nu + N$  \\
(b) $\bar\nu_e + p \leftrightarrow n + e^+$        &   \hspace{0.8cm} (h) $\nu + A \rightarrow \nu + A$   \\
(c) $\nu_\mu + n \leftrightarrow p + \mu^-$                 &   \hspace{0.8cm} (i) $e^- + e^+ \leftrightarrow \nu + \bar\nu$  \\
(d) $\bar\nu_\mu + p \leftrightarrow n + \mu^+$       &  \hspace{0.8cm} (j) $N + N \leftrightarrow  N+N +\nu + \bar\nu$ \\
(e) $e^- + p + \bar\nu_e \leftrightarrow n$  &   \hspace{0.8cm} (k) $\mu^- \leftrightarrow e^- + \nu_\mu + \bar\nu_e$\\
(f) $\mu^- + p + \bar\nu_\mu \leftrightarrow n$ &  \hspace{0.8cm} (l) $\mu^+ \leftrightarrow e^+ + \bar\nu_\mu + \nu_e$
\\
\hline
\end{tabularx}
    \label{tab:reactions}
\end{table}

\section{Results}

\begin{figure*}[htpb!]
    \centering
    \begin{subfigure}{}
    \includegraphics[width=\linewidth]{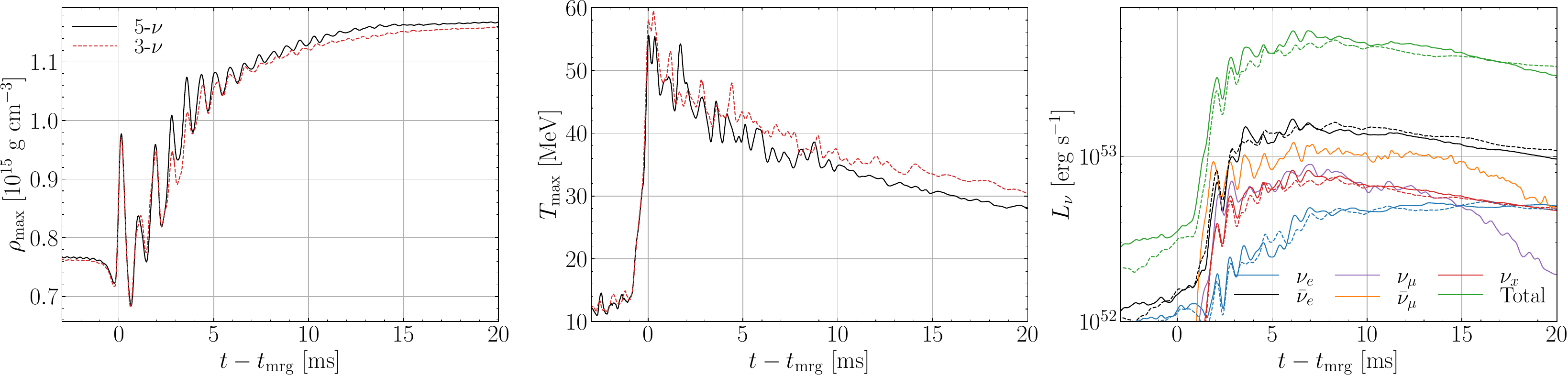}
    \end{subfigure}
    \begin{subfigure}{}
    \includegraphics[width=\linewidth]{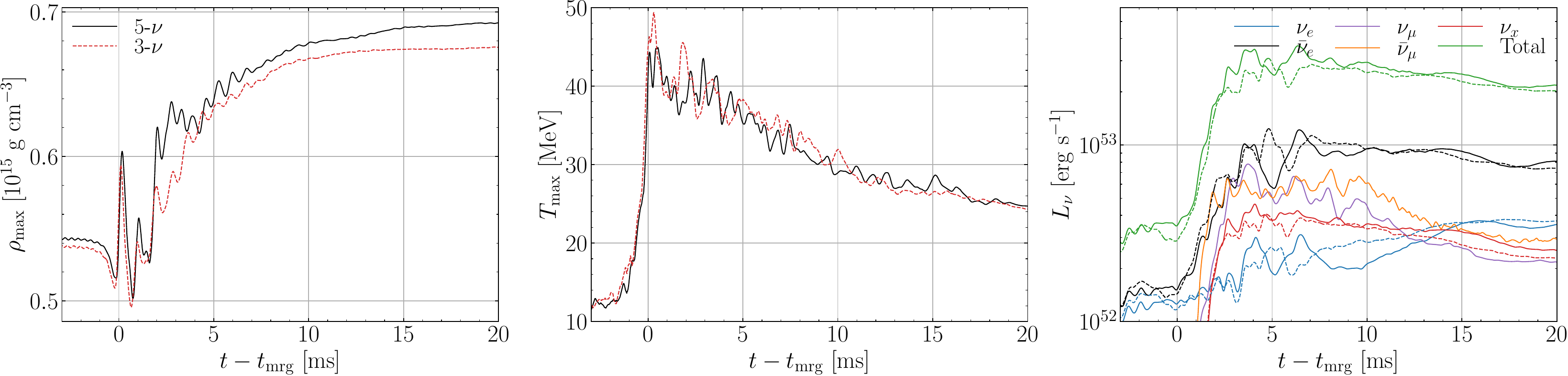}
    \end{subfigure}
    \caption{Evolution of diagnostic quantities for the SFHo (upper panels) and DD2 (lower panels) runs. For a better visualization, a moving-average with window size of $0.2~{\rm ms}$ is applied. Left column: maximum rest-mass density for the $5$-$\nu$ (black thick line) and $3$-$\nu$ (red dashed line). Middle column: maximum temperature. Right column: neutrino luminosities of all species for the $3$ (dashed lines) and $5$-$\nu$ (thick lines) cases. Luminosities are computed as in (\cite{Schianchi:2023uky}) on a coordinate sphere of fixed radius $r \approx 300~{\rm km}$. For a clearer comparison, the luminosity of heavy-lepton neutrinos $\nu_x$ is divided by 2 (4) for $5$-$\nu$ ($3$-$\nu$).}\label{fig:diags}
\end{figure*}

The initial data for our simulations, constructed using the \texttt{SGRID} code (\cite{Tichy:2019ouu}), represent mass-symmetric systems with individual masses in isolation of $M = 1.25~M_\odot$ employing a cold $T = 0.1~{\rm MeV}$ and in neutrinoless $\beta$-equilibrium slice of the electronic/muonic EOSs. The initial coordinate distance is $d_0 \approx 41.4~{\rm km}$, yielding $\sim 4$ ($\sim 5$) full orbits before merger for the DD2 (SFHo) setups, regardless of the presence of muons. For convenience, we present results shifted in time with respect to the merger time $t_{\rm mrg}$, by the moment when the dominant $(2,2)$ mode of the GW signal reaches its maximum amplitude. Evolution of diagnostic quantities are shown for our SFHo and DD2 runs in Figure~\ref{fig:diags}. In the left column we see larger maximum densities for the $5$-$\nu$ (thick black lines) compared to the $3$-$\nu$ case (red dashed lines), both before and after merger, caused by the softening of the EOS in the presence of muons. At later times $t-t_{\rm mrg} \gtrsim 10~{\rm ms}$, when the internal motions are dampened by emission of GWs, the $5$-$\nu$ remnants and their $3$-$\nu$ counterparts become similarly dense, with differences of $\sim 1\%$ ($\sim 3\%$) for SFHo (DD2). In fact, the larger softening seen in DD2 $5$-$\nu$ is due to the  $\sim 15\%$ higher central $Y_\mu$ (and hence smaller neutron fraction) compared to SFHo $5$-$\nu$, which is set by the initial data. Shortly after the merger, comparable muonization in both baselines drives $Y_\mu$ higher for DD2, which reduces pressure support from degenerate neutrons, as will become clearer in the following. In the middle column we observe moderately smaller maximum temperatures for SFHo $5$-$\nu$ during the merger aftermath, attributed to additional cooling mechanisms via muonic reactions, while for DD2 $5$-$\nu$, temperatures are much closer to its $3$-$\nu$ counterpart. Then, in the right column we depict the neutrino luminosities, where similar evolutions are reported for $\nu_e$, $\bar\nu_e$ and $\nu_x$ in all cases, while contributions from $\bar\nu_\mu$ and $\nu_\mu$ are such that, for $5$-$\nu$ cases, the combined luminosity $2L_{\nu_x} + L_{\nu_\mu} + L_{\bar\nu_\mu}$ is larger than $4L_{\nu_x}$ in the $3$-$\nu$ case. By the end of our simulations (not shown), combined luminosities of $\{\nu_\mu,\bar\nu_\mu,\nu_x\}$ indeed converge to the same value as $\nu_x$ in the $3$-$\nu$ counterparts.
Furthermore, $L_{\bar\nu_e} > L_{\nu_e}$ in all scenarios indicates electronization of matter.

\begin{figure*}[htpb!]
    \centering
    \begin{subfigure}{}
    \includegraphics[width=\linewidth]{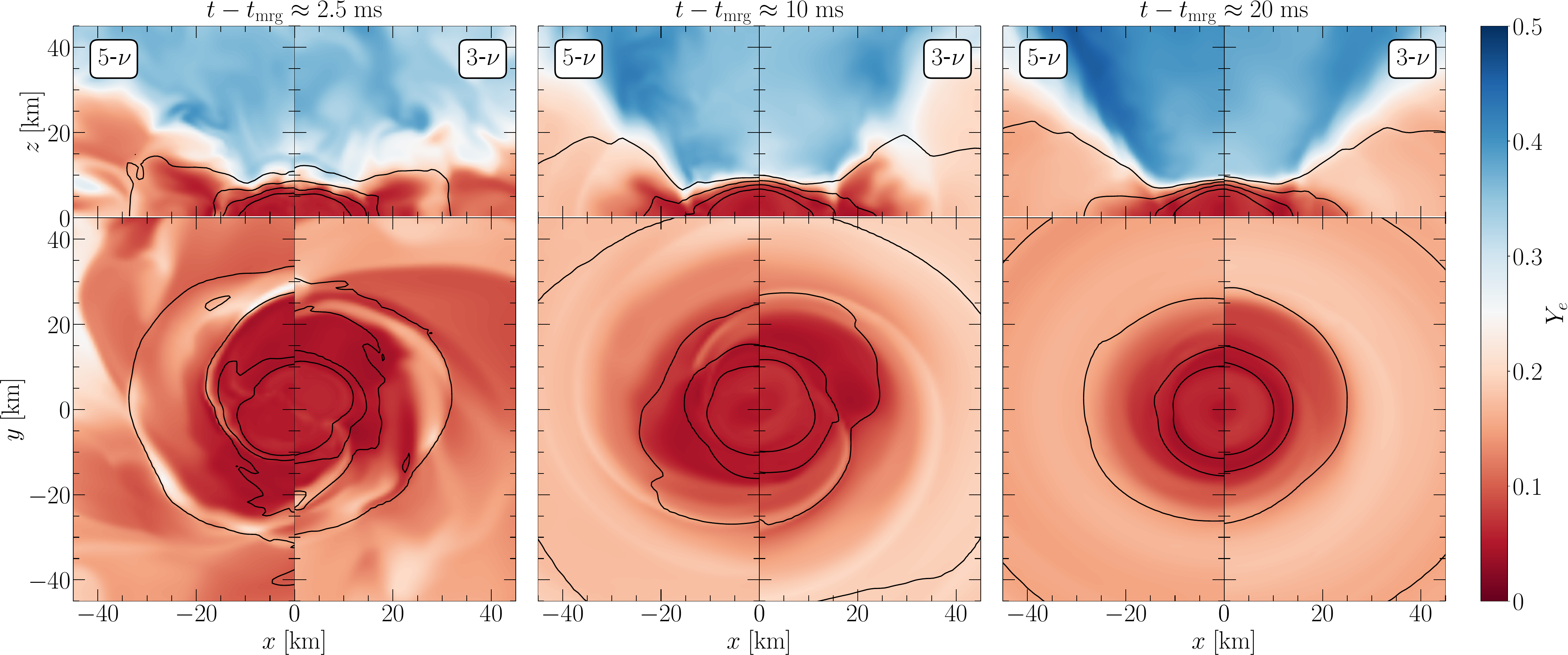}
    \end{subfigure}{}
    \begin{subfigure}{}
    \includegraphics[width=\linewidth]{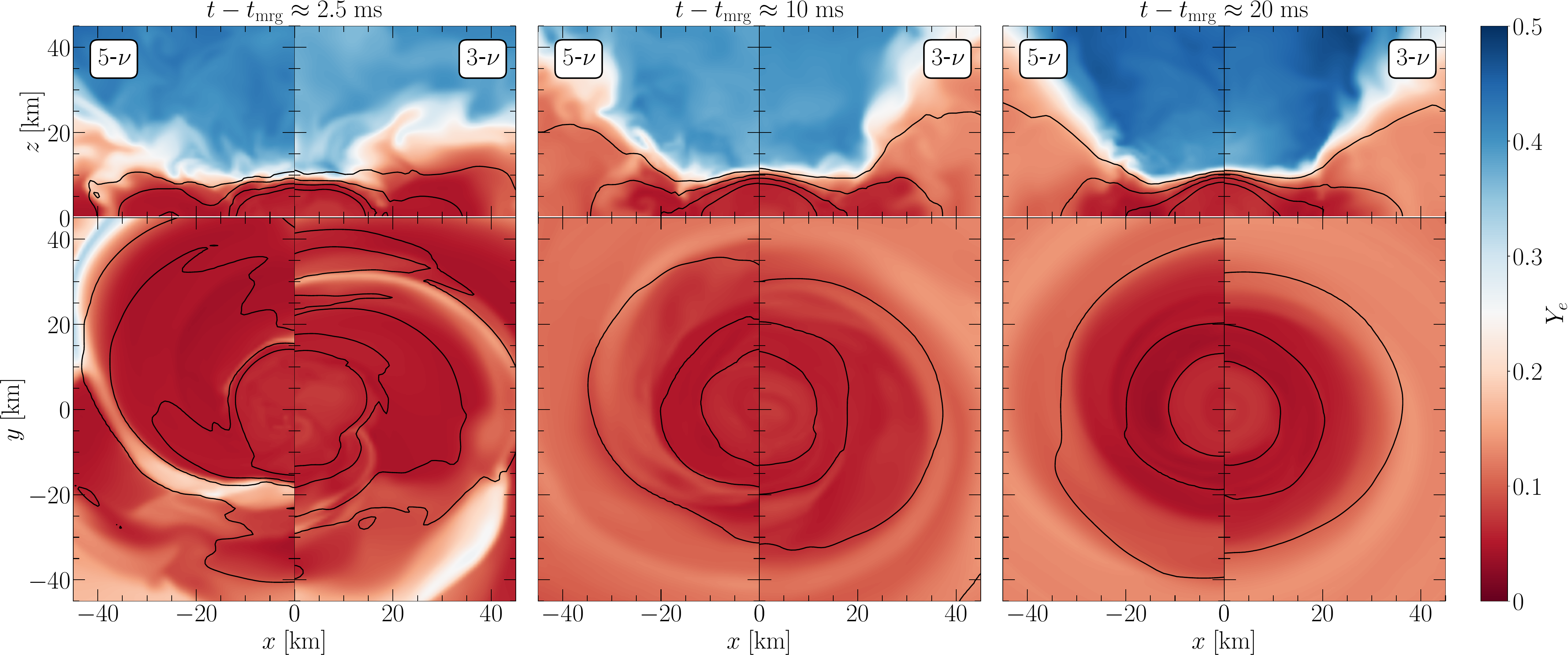}
    \end{subfigure}{}
 \caption{Snapshots of the electron fraction for the SFHo (upper panels) and DD2 (lower panels) in the polar plane and orbital plane. The left (right) half represents $5$-$\nu$ ($3$-$\nu$) runs at $t-t_{\rm mrg} = \{2.5, 10, 20\}~{\rm ms}$ from left to right columns. Black lines are isodensity contours from $\{10^{11},10^{12}, 10^{13}, 10^{14}\} ~{\rm g~cm^{-3}}$. Regardless of the baryonic baseline, in both the muonic and electronic runs we note early electronizaton of matter at low densities (left and middle columns), as well as the beginning of the secular ejection of highly electronized material around the pole (right column) via neutrino winds.}
 \label{fig:ye-snp}
\end{figure*}

As shown in Figure~\ref{fig:ye-snp}, we note qualitatively similar $Y_e$ distributions throughout the post-merger. For instance, early electronization of matter (left column) at densities smaller than $10^{12}~{\rm g~cm^{-3}}$ is caused by excess free emission of $\bar\nu_e$ compared to $\nu_e$, as the former decouples from matter deeper in the remnant than the latter. Early electronization is marked by $Y_e \gtrsim 0.25$ at low densities both in the disk (lower portion of the snapshots), where matter obstructs the propagation of neutrinos coming from the remnant, and in the polar cap (upper portion of the snapshots), where further electronization is promoted by a dominant absorption of $\nu_e$. This effect is better observed in the middle and right columns as a gradual increase in $Y_e$ along a funnel centered at the remnant. 

Next, we address the evolution of the muonic content for our $5$-$\nu$ simulations via a few diagnostic quantities. First is the total muon number
\begin{equation}\label{eq:cons-mu}
    N_{\mu} = \frac{1}{m_b}\int_{\mathcal{V}} d^3x~\sqrt{\gamma} W\rho Y_\mu,
\end{equation}
where the integration volume $\mathcal{V}$ hereafter will be the box corresponding to the second coarsest grid level. Second, useful to analyze the evolution of leptonic species $l$ in the remnant is the total lepton number fraction. For muons, it is given by
\begin{equation}\label{eq:ylm}
    Y_{L\mu} = Y_\mu + Y_{\nu_\mu} - Y_{\bar\nu_\mu},
\end{equation}
and encompasses the net muon fraction in matter $Y_\mu$ and the fraction of muon (anti)neutrinos ($Y_{\bar\nu_\mu}$) $Y_{\nu_\mu}$, given by
\begin{equation}
    Y_{\nu/\bar\nu} = \frac{n_{\nu/\bar\nu}}{n_b}.
\end{equation}
Eq.~\eqref{eq:ylm} is useful because in BNS merger remnants, neutrinos are expected to be trapped and achieve chemical/thermal equilibrium with the fluid, at least in sufficiently hot matter (\cite{Alford:2026kwd}). Such a mixture of matter and radiation in equilibrium is described by an EOS parameterized by $(n_b, T, Y_p = Y_e + Y_\mu,Y_{L\mu})$, such that the chemical potential of muons can be determined by solving Eq.~\eqref{eq:eq-ym-5} for each thermodynamic point of the baryonic baseline over a range of $Y_{L\mu}$. As a result, simulation data can be used to infer the expected muon fraction $\tilde{Y}_{\mu} = \tilde{Y}_{\mu}(n_b, T, Y_p, Y_{L\mu})$ when matter and radiation equilibrate. With this construction, the connection between leptonic composition of matter and microscopic properties of the baryonic baseline is more evident, since the difference $\mu_n - \mu_p$ entering the equilibrium chemical potential of neutrinos correlates to the baryonic symmetry energy~(\cite{Most:2021ktk, Gieg:2025ivb}).

Lastly, we define the relative deviation of net muon fraction from equilibrium as
\begin{equation}
    \frac{\delta Y_\mu}{Y_\mu} = 1 - \frac{\tilde{Y}_\mu(n_b, T, Y_p,Y_{L\mu})}{Y_\mu},
\end{equation}
and its mass-average in the orbital plane as
\begin{equation}\label{eq:avg-dev}
    \left\langle \frac{\delta Y_\mu}{Y_\mu} \right\rangle_{xy} = \frac{ \displaystyle{\int} dx dy~\sqrt{\gamma} W \rho \frac{\delta Y_\mu}{Y_\mu}}{\displaystyle{\int} dxdy~\sqrt{\gamma} W\rho},
\end{equation}
where the integration is carried out in the second finest level, containing the remnant and a small portion of the disk.

\begin{figure*}[htpb!]
    \centering
    \begin{subfigure}{}
            \includegraphics[width=\linewidth]{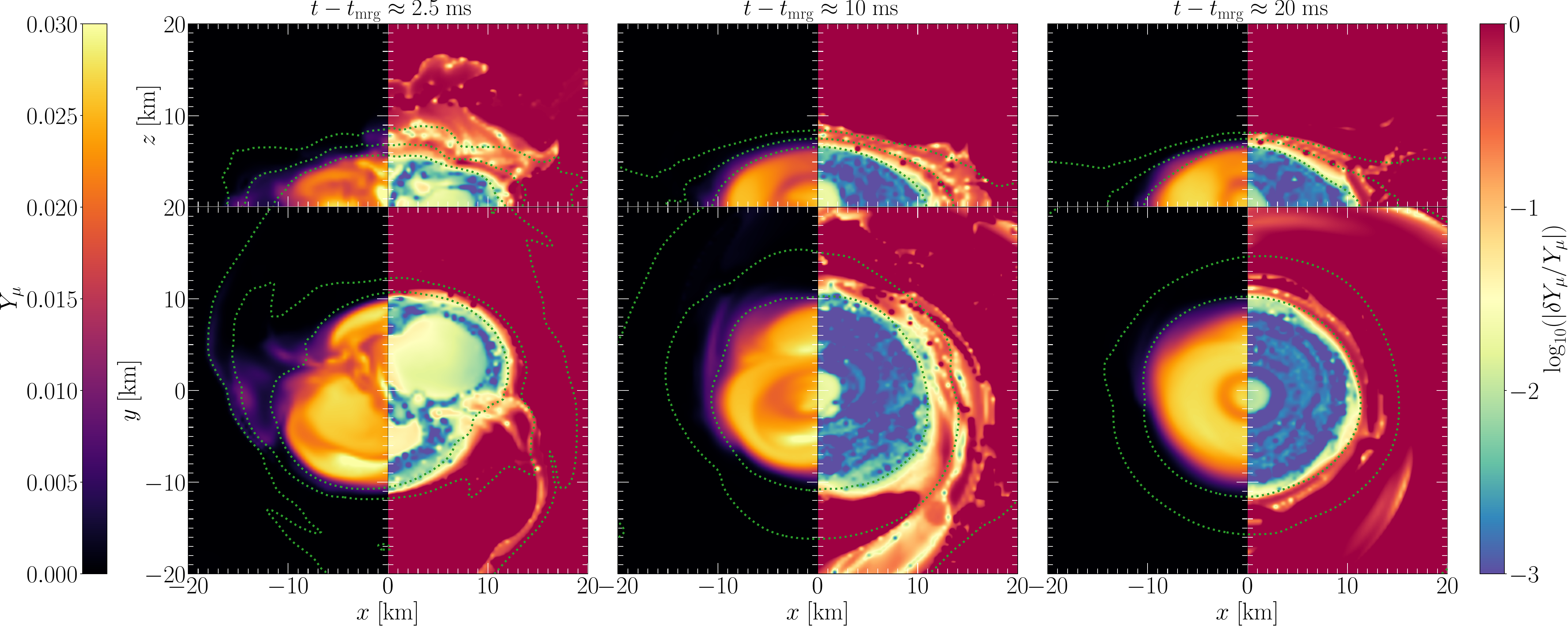}   
    \end{subfigure}{}
    \begin{subfigure}{}
            \includegraphics[width=\linewidth]{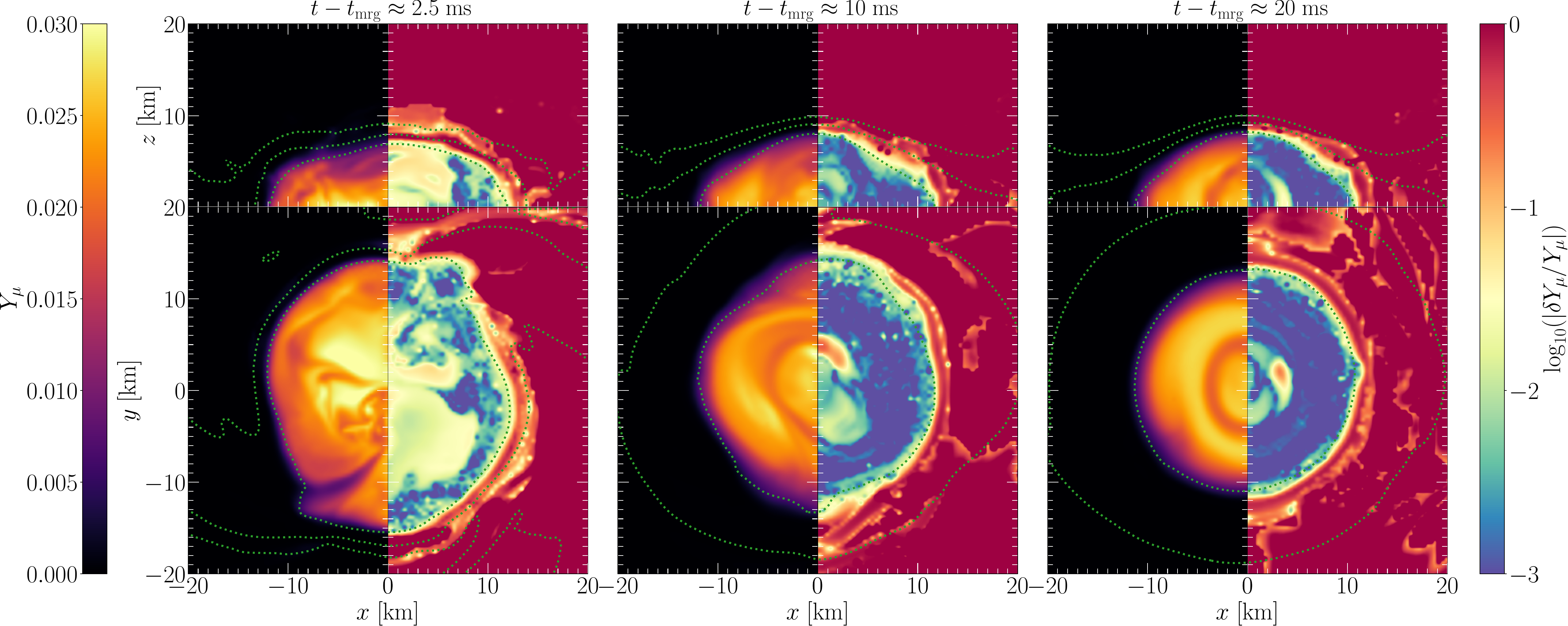}   
    \end{subfigure}{}
 \caption{Snapshots of the muon fraction $Y_\mu$ (left half) and relative difference with respect to equilibrium $|\delta Y_\mu/Y_\mu|$, Eq.\eqref{eq:avg-dev}, (right half) in the polar plane and orbital plane for the SFHo $5$-$\nu$ (upper panels) and DD2 $5$-$\nu$ (lower panels) runs at $t-t_{\rm mrg} = \{2.5, 10, 20\}~{\rm ms}$ from left to right columns. Green-dotted lines mark constant densities $\{10^{12}, 10^{13},10^{14}\}~{\rm g~cm^{-3}}$.}\label{fig:ym-snp}
\end{figure*}

In Figure \ref{fig:ym-snp}, we present the post-merger profiles of net muon fraction $Y_\mu$ in the left half and the relative deviation from equilibrium in the right half of each snapshot for SFHo $5$-$\nu$ (upper panels) and DD2 $5$-$\nu$ (lower panels). Note that the smaller the deviation (yellow to blue), the better the description of a mixture of matter/radiation in equilibrium. Hence, there (anti)neutrinos are trapped. Larger deviations (red), on the other hand, correspond to out-of-equilibrium states, where (anti)neutrinos are not trapped.
Regardless of the baryonic baseline, we note that largest deviations from equilibrium are located at $\rho \lesssim 10^{13}~{\rm g~cm^{-3}}$ in the early post-merger (left columns), while the smallest deviations $\sim 10^{-3}$ are found in hotter regions, e.g., the shear layer and the hot outskirts close to $\rho \sim 10^{14}~{\rm g~cm^{-3}}$. At $t - t_{\rm mrg} \approx 10~{\rm ms}$ (middle column), as the remnant cools, larger portions of the interior $\rho \gtrsim 10^{14}~{\rm g~cm^{-3}}$ equilibrate with deviations of a few percent or less, except for DD2 (lower panels), where deviations of $\sim 30 \%$ persist in a localized spot around the center. At $t-t_{\rm mrg} \approx 20~{\rm ms}$ (right column), the innermost portions of the remnants achieve equilibrium, evidencing that our equilibration prescription accurately captures the formation of a trapped gas of (anti)neutrinos.

It is interesting to note that the bulk of the net muon fraction is found at $\rho \geq 10^{14}~{\rm g~cm^{-3}}$, while some $Y_\mu \sim 0.010 -0.015$ is present at smaller densities, but hotter matter streams, especially notable in the left column for SFHo due to its higher temperatures. In view of 
our deviation from equilibrium argument, we observe that the trapping of muonic neutrinos require substantial muon fractions $Y_\mu \sim 0.020$, for which muonic $\beta$-reaction rates are important.

On the contrary, at smaller densities, larger deviations suggest that muon (anti)neutrinos are on a (partially) free-streaming regime, where equilibrium states of matter approach a neutrinoless condition~(\cite{Espino:2023dei}). Therefore, negligible muon fractions are to be expected (see Figure 14 of \cite{Gieg:2024jxs}), because muonic $\beta$-reactions drive matter with $\rho \leq 10^{14}~{\rm g~cm^{-3}}$ to neutrinoless $\beta$-equilibrated states with $Y_\mu \ll10^{-3}$. Higher muon fractions at lower densities are unlikely, requiring temperatures $T \gtrsim 15~{\rm MeV}$.

\begin{figure}[htpb!]
    \centering
    \includegraphics[width=\linewidth]{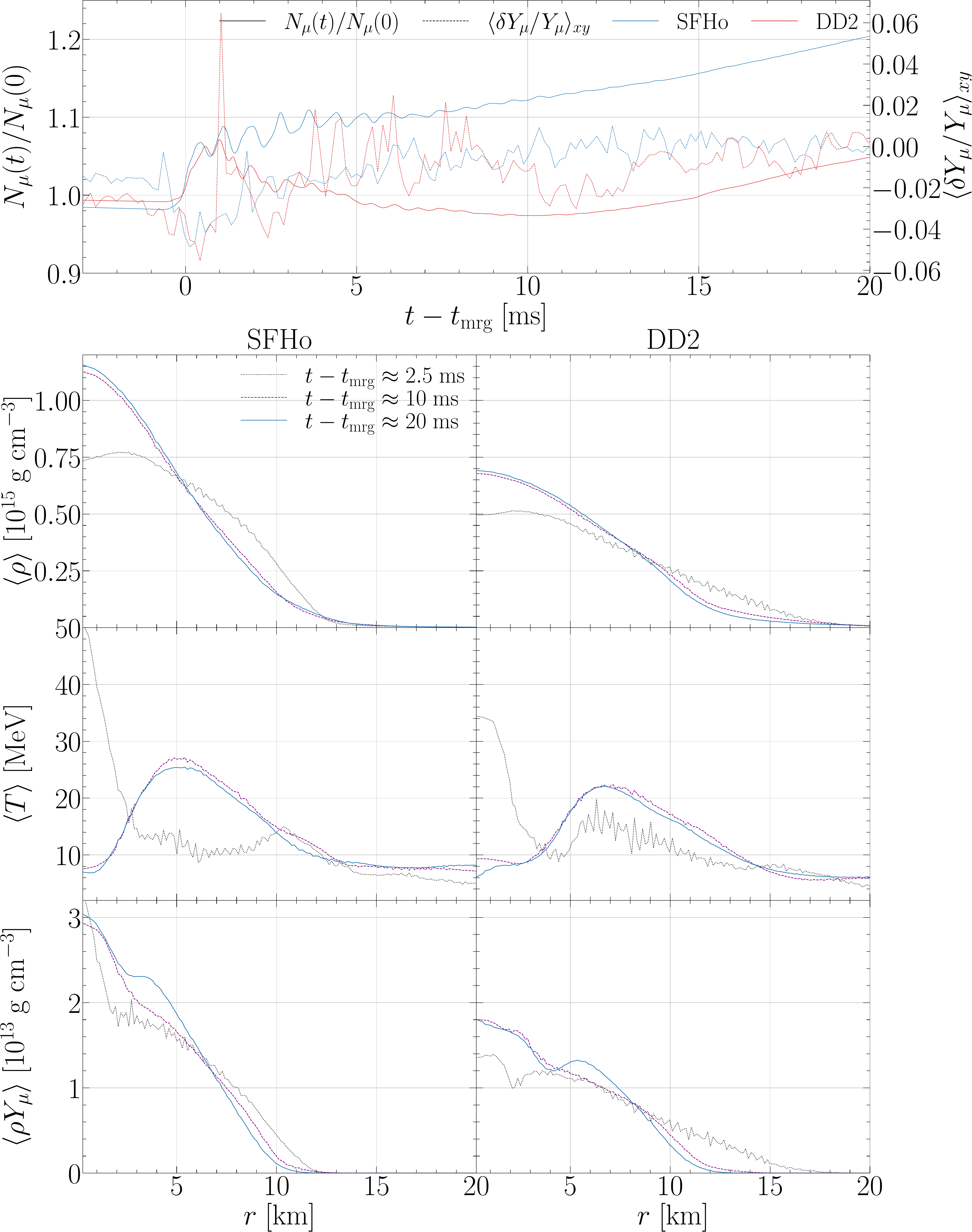}
    \caption{Diagnostics for the muonic content. Upper panel: evolution of normalized total muon number (thick lines) and mass-averaged deviation from equilibrium (dotted) for our $5$-$\nu$ runs SFHo (blue) and DD2 (red). Lower panels: from top to bottom, we present radial averages of rest-mass density, temperature and muon density at different times along the post-merger for SFHo (left column) and DD2 (right column).}
    \label{fig:dym-deltaym}
\end{figure}

For a more quantitative view of the dynamics of the muonic content, we present in the upper panel of Figure~\ref{fig:dym-deltaym} the evolution of the conserved muon number, Eq.~\eqref{eq:cons-mu}, normalized to its initial value (left axis, thick lines), and the mass-averaged deviation from equilibrium,  Eq.\eqref{eq:avg-dev}, (right axis, dotted lines) for the muonic SFHo (blue) and DD2 (red) runs. There we observe a reasonable $1-2\%$ conservation of total muon number along the inspiral, with a steep increase during the first millisecond of the post-merger, followed by a steady increase of $N_\mu$ (blue, thick line) for SFHo up to $\sim 20\%$ at $t-t_{\rm mrg} \approx 20~{\rm ms}$. The average deviation (blue, dotted line) stays at $\sim -1.2\%$ before merger, with larger deviation up to $\sim -4\%$ that gradually converges to $\sim 0$ as the remnant transitions to a compositional equilibrium. For DD2, $N_\mu$ (red, thick line) decreases after the initial burst up to $t-t_{\rm mrg} = 10~{\rm ms}$, and then increases to $\sim 4-6\%$ by the end. In this case, noticeably larger deviations (red, dotted line) compared to SFHo are sustained, only to converge to $\sim 0$ over a longer timescale. Since both runs converge towards small deviation, we conclude that the transition from non-equilibrium to equilibrium state is captured, although evolution of the muonic content shows no clear trend besides late time muonization, while details of the transitional regime depend on the baryonic model in a non-trivial manner. To avoid confusion, it is important to clarify that our equilibrium analysis is based on the local equilibrium states of matter and radiation within an environment that is dynamically cooling and compacting. As such, the equilibrium configurations change over time. That is why we may still have ongoing muonization on the remnant, while diminishing deviations indicate matter and radiation achieving local equilibrium. Longer simulations would be required to observe whether the conserved muon number stabilizes.

For a complementary view on how matter evolves and the locations of muonization in the remnant, we present in the lower panels of Figure~\ref{fig:dym-deltaym}, respectively from top to bottom, averaged radial profiles of rest-mass density, temperature and muon density at selected instants. Note that we depict the muon density $\rho Y_\mu$, because this quantity embodies density variations, making the loci of muonization clearer. For SFHo (left column), we observe substantial early (black, dashed line) muonization on a narrow $r \leq 2~{\rm km}$, coinciding with the high temperatures of the shear layer. At $t - t_{\rm mrg} \approx 10~{\rm ms}$ (purple, dotted lines), the remnant becomes denser and colder at the center, the hot annulus is formed at $2 \leq r \leq 12~{\rm km}$, and the bulk of muons redistributes to a larger radius $r \leq 5~{\rm km}$. Then, at $t - t_{\rm mrg}\approx 20~{\rm ms}$ the remnant contracts and cools slightly, while a layer $2.5 \leq r \leq 4.5~{\rm km}$ of higher muon density arises. Hence, we identify the main regions of muonization as the center and within the hot annulus, not exactly at the temperature peak (where $T\sim 24~{\rm MeV},~\rho \sim 7\times10^{14}~{\rm g~cm^{-3}}$), but at slightly smaller temperature and higher $\rho \sim 8.5\times10^{14}~{\rm g~cm^{-3}}$. For DD2 (right column), we observe at $t-t_{\rm mrg}\approx 2.5~{\rm ms}$ smaller muon densities in the shear layer $r \leq 3~{\rm km}$. Similar to SFHo, at $t-t_{\rm mrg} \approx 10~{\rm ms}$ we see the formation of a hot annulus at $3 \leq r \leq 14~{\rm km}$ and muonization of inner layers $r \leq 4~{\rm km}$, while, in contrast, at $t - t_{\rm mrg} \approx 20~{\rm ms}$, muonization is more evident within the hot annulus than in the core. Furthermore, we remark that muonization is dominated by muon neutrino processes, rather than muon antineutrinos. In particular, absorption of $\nu_\mu$ on $n$ generally decreases fluid thermal energy, as the creation of $\mu^-$ converts part of the fluid energy to rest-mass $m_\mu \sim 106~{\rm MeV}$. On the other hand, inverse muon decay deposits thermal energy in the fluid, as it converts energy from pairs $\nu_\mu,~\bar\nu_e$ and $e^-$ into $\mu^-$. For example, a pair of thermal neutrinos and a non-degenerate, ultrarelativistic electron at $T \sim 30~{\rm MeV}$ would contribute $\sim 3T = 90~{\rm MeV}$ each. The resulting muon would, then, deposit $\sim 164~{\rm MeV}$ of kinetic energy in the fluid, larger than the kinetic energy from the electron. This phenomenology plausibly explains the inversion in maximum temperatures for DD2 $5$-$\nu$ and its $3$-$\nu$ counterpart at $t-t_{\rm mrg}\geq 10~{\rm ms}$ (see Figure~\ref{fig:diags}), which coincides with the gradual muonization depicted in the upper panel of Figure~\ref{fig:dym-deltaym}. Hence, the interpretation that our reported cooling is less potent than that of \cite{Ng:2024zve} due to inverse muon decay is tempting.

We also highlight the good agreement of the $Y_\mu$ distributions found at $t - t_{\rm mrg} \approx 20~{\rm ms}$ and those reported in Figure 8 of the core-collapse supernovae simulations of \cite{Fischer:2020vie}, i.e., peaks of $Y_\mu$ are located preferentially at high densities (remnant's center) and high temperatures (hot annulus).

Finally, we discuss the remnant and ejecta properties. For that, material is considered ejecta if, at a fixed coordinate sphere of radius $r \approx 300~{\rm km}$, the matter element satisfies the geodesic criterion (Eq. (13) of \cite{Dietrich:2015iva}. On the contrary, bound matter is considered to be part of the remnant if $\rho \geq 10^{13}~{\rm g~cm^{-3}}$, while parts of the disk have $\rho < 10^{13}~{\rm g~cm^{-3}}$. Relevant data is summarized in Table~\ref{tab:sum-prop}. There we note good agreement for the disk and remnant baryonic masses for the $3$-$\nu$ runs and their respective $5$-$\nu$ counterparts, as well as average thermodynamical properties of the ejecta, such as average electron fraction $\langle Y_e\rangle_{\rm ej}$, average asymptotic velocity $\langle v_\infty\rangle_{\rm ej}$ and average temperature $\langle T \rangle_{\rm ej}$. As anticipated, the addition of muonic reactions as energy sink for outflowing material results in overall smaller ejecta for the $5$-$\nu$ runs, about $\sim 17\%$ less, instead of the $50\%$ reported in our previous leakage simulations (\cite{Gieg:2024jxs}). This is because here we account for ejection via absorption of neutrinos. Also, for 5-$\nu$, the average electron fraction in the remnant gets smaller, but charge neutrality enforces higher proton fractions than in the $3$-$\nu$ case.

\begin{table*}[t]
    \centering
    \caption{Summary of remnant, disk and ejecta properties at $t - t_{\rm mrg} \approx 20~{\rm ms}$. From left to right columns read simulation name, baryonic mass of the disk, baryonic mass of the remnant, mass-averaged electron fraction in the remnant, mass-averaged muon fraction in the remnant (only for the $5$-$\nu$), ejecta mass, average electron fraction of the ejecta, average asymptotic velocity of the ejecta and average temperature of the ejecta.}
\begin{tabular}{cccccccccc}
\hline
Simulation & $M_{\rm disk}~[M_\odot]$ & $M_{\rm rem}~[M_\odot]$ &$\langle Y_e\rangle_{\rm rem}$  & $\langle Y_\mu\rangle_{\rm rem}$  & $M_{\rm ej}~[10^{-3} M_\odot]$ & $\langle Y_e\rangle_{\rm ej}$ & $\langle v_\infty\rangle_{\rm ej}$ & $\langle T\rangle_{\rm ej}~[{\rm MeV}]$ \\
SFHo $3$-$\nu$ & $0.104$ & $2.59$ & $0.064$ & - & $2.16$ & $0.37$ & $0.17$ & $0.64$ \\
SFHo $5$-$\nu$ & $0.109$ & $2.58$ & $0.057$ & $0.023$ & $1.85$ & $0.35$ & $0.17$ & $0.63$ \\
DD2 $3$-$\nu$ & $0.178$ & $2.46$ & $0.066$ & - &$1.91$ & $0.32$ & $0.14$ & $0.63$  \\
DD2 $5$-$\nu$ & $0.197$ & $2.41$ & $0.060$ & $0.020$ &$1.68$ & $0.31$ & $0.14$ & $0.64$ \\
\hline
\end{tabular}
    \label{tab:sum-prop}
\end{table*}

\begin{figure*}[htpb!]
    \centering
    \begin{subfigure}{}
            \includegraphics[width=\linewidth]{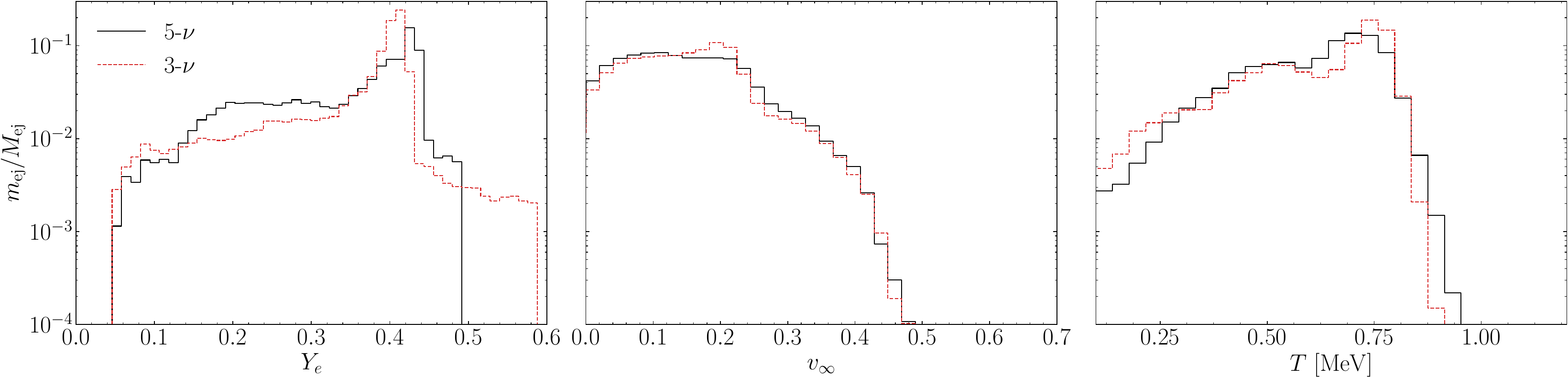}   
    \end{subfigure}{}
    \begin{subfigure}{}
            \includegraphics[width=\linewidth]{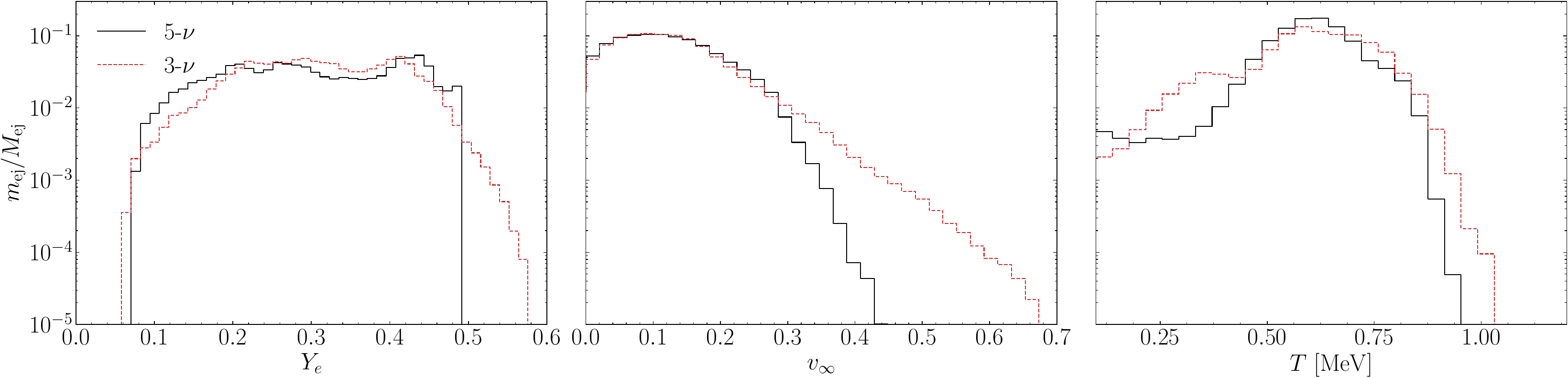}   
    \end{subfigure}{}
 \caption{Mass-normalized distributions of ejecta properties up to $t - t_{\rm mrg} \approx 20~{\rm ms}$: electron fraction (left column), asymptotic velocity (middle column) and temperature (right column) for our SFHo (upper panels) and DD2 (lower panels) runs. Note that at the extraction sphere ($r \approx 300~{\rm km}$), and there $Y_\mu = 10^{-4}$.}
 \label{fig:hists}
\end{figure*}

More details about the ejecta properties are provided in Figure~\ref{fig:hists}. The ejecta properties for $3$ and $5$-$\nu$ simulations are in good qualitative agreement. On a quantitative level, from the left column, upper panel, we observe for SFHo $5$-$\nu$ larger amounts of material with $Y_e \sim 0.14 - 0.30$ compared to its $3$-$\nu$ counterpart. The peak in both cases lies in $Y_e \sim 0.42 - 0.44$, with a larger peak fraction in the $3$-$\nu$ case, suggesting stronger neutrino irradiation on the ejecta, consistent with the higher temperatures for this case. We also remark that the sharp drop in $Y_e = 0.5$ for the muonic runs is caused by the EOS limit to this value. Hence, it is plausible that higher values could be achieved. No significant deviations are noted for the asymptotic velocity (middle column, upper panel) and temperature (right column, upper panel). For our DD2 runs (lower panels), we see larger fractions of material with $Y_e \sim 0.1 - 0.2$ in the $5$-$\nu$ run, and the peak is also located at slightly higher $Y_e$ compared to $3$-$\nu$. 
Most of the differences are seen in the asymptotic velocity, where no material with $v_\infty \gtrsim 0.4 c$ is present, while the $3$-$\nu$ run steadily extends to $v_\infty \approx 0.6c$. To explain this observation, we note that the difference in total luminosity between $5$-$\nu$ and $3$-$\nu$ is larger for DD2 than for SFHo (see Figure~\ref{fig:diags}) during the first $\sim 5~{\rm ms}$ of the post-merger. On the other hand, fast ejecta is produced within this timescale, mainly arising from the shear layer and in the first bounces of the merging cores~(\cite{Rosswog:2024vfe}). Thus, we assume that this ejecta component is subject to stronger radiative losses, leading to smaller fractions of fast outflow. Interestingly, the shifted peaks of $Y_e$ towards higher values for both muonic runs result from the decay of muons at densities $\rho \leq 10^{12}~{\rm g~cm^{-3}}$ (not shown).

\section{Conclusions}
In this \textit{Letter} we presented a set of BNS merger simulations employing state-of-art neutrino interaction rates, computed within the full kinematics approach, and a more accurate M1 neutrino transport scheme using an implicit-explicit time integrator. We assessed the possible imprints of muons and muonic weak interactions in the properties of the remnant, disk and ejected material, the latter closely related to nucleosynthetic yields and kilonovae signals. Overall, comparing our $3$ and $5$-$\nu$ simulations for two baryonic baseline EOSs (SFHo and DD2), we see modest impacts in the equilibrium structure of the remnant, with slightly denser cores (at most $3\%$), slightly cooler core-disk interface (by at most $2~{\rm MeV}$), and comparable neutrino luminosities. Accordingly, the post-merger compositional and thermal evolution, respectively given by $Y_e$ and $T$, is not very sensitive to the presence of muons. Indeed, we were able to capture the equilibration process of matter and radiation with our novel two timescales approach, and showed that the presence of muons is restricted to densities in excess of $\rho = 10^{14}~{\rm g~cm^{-3}}$. On the macroscopic side, we showed that disk and remnant masses are broadly consistent in both scenarios (within $\sim 5\%$), and that ejecta masses are smaller by less than $\sim 17\%$ for muonic runs, while average electron fraction, asymptotic velocity and temperature of the outflows are consistent within $\sim 5\%$. Therefore, our main conclusion is that neglecting the presence of muons has limited impacts in the outcomes of a BNS merger simulation, at least for small mass, symmetric systems such as those considered in this work. Consequently, we expect that nucleosynthetic yields and the associated kilonova signal to not be significantly modified by the presence of muons and muonic weak reactions. 

We note, however, important contrasts between our results and those presented in \cite{Ng:2024zve}, where it is reported much larger fractions ($\sim 20 - 30\%$) of low $Y_e \lesssim 0.05$ ejecta, with almost absent outflows with $Y_e \geq 0.3$ in the muonic case, which leads to larger abundances of lanthanides, but smaller abundances of lighter elements. This is mainly due to their choice of methods for handling the possibly large negative muon neutrino degeneracies at low densities. A detailed analysis of methodological differences that could account for the discrepancies is presented in Sec.~\ref{sec:lit-comp} of the Supplementary Material.

That said, we highlight the importance of a consistent treatment of neutrino interactions for accurately capturing the compositional evolution of the system. Our results suggest that a kernel-based approach for the computation of pair-processes interaction rates, albeit approximated by the assumption of LTE for pairing neutrinos, is a suitable solution to the divergent behavior of pair opacities for (anti)muon neutrinos, offering the advantage of a unified treatment of weak rates for all neutrino species. Future work with more elaborate transport schemes, such as spectral M1 or Monte-Carlo based methods~(\cite{Kawaguchi:2024naa, Kawaguchi:2025con, Foucart:2022kon, Foucart:2024npn}) would quantify the uncertainties introduced by our present approach. For future investigations we envision plenty of options, e.g., the role of magnetic fields in muonic BNS mergers; exploration of wider regions of the BNS parameter space (different mass-asymmetries, higher total masses, spinning configurations); the impact of $5$-$\nu$ treatments with neutrino oscillations (\cite{Qiu:2025kgy, Qiu:2025ybw}); or the use of more complete microphysics, such as inclusion of pions in the EOS (\cite{Vijayan:2023qrt}) and their reactions with neutrinos, specially relevant for muon (anti)neutrinos (\cite{Fore:2019wib}) in hot matter.

\section{Acknowledgments}
It is a pleasure to thank Guilherme Grams and Harry Ho-Yin Ng for fruitful discussions. HG, RJ and TD acknowledge funding from the EU Horizon under ERC Starting Grant, no. SMArt101076369. MU acknowledge S\~ao Paulo Research Foundation (FAPESP), Brasil, for financial support under Process No. 2024/21086-5. 
Views and opinions expressed are those of the authors only and do not necessarily reflect those of the European Union or the European Research Council. Neither the European Union nor the granting authority can be held responsible for them. 
The simulations were performed on the HPC systems Lise/Emmy of the North German Supercomputing Alliance (HLRN) [project bbp00049], and on the DFG-funded research cluster Jarvis at the University of Potsdam (INST 336/173-1; project number: 502227537).

\bibliography{references}

\appendix
\section{Neutrino Rates}\label{sec:nu-rates}
The weak reactions considered in this work are summarized in Table~\ref{tab:reactions}. Spectral (energy-dependent) opacities and annihilation kernels are produced for all thermodynamical points $(\rho$, $T$, $Y_e$, $Y_\mu)$ in the EOS range at neutrino energies $\epsilon = [0.5,420]~{\rm MeV}$, binned in 18 $\log_{10}$ spaced intervals. For the $\beta$-reactions (a) -- (f) in Table~\ref{tab:reactions}, we follow (\cite{Guo:2020tgx}) in the full kinematics approach, i.e., considering self-consistent modifications to relativistic hadronic currents due to mean-field potentials and effective masses, weak-magnetism, pseudoscalar terms, and momentum-dependent nucleon form factors. The integration over the phase-space of hadrons is split in $\epsilon$. For $\epsilon \leq 120~{\rm MeV}$, expansion of coupling constants at low momentum exchange are accurate, allowing opacities to be evaluated by 2-dimensional numerical integrals over energies. For $\epsilon \geq 120~{\rm MeV}$, such an expansion is inaccurate and we numerically perform 4-dimensional integrals over energy, momentum exchange and two angles. The corresponding spectral opacities are, then, corrected for stimulated absorption and averaged with respect to the equilibrium Fermi-Dirac distribution to produce gray opacities.
Iso-energetic scattering on free nucleons (g) receives energy-dependent corrections for weak magnetism/recoil (\cite{Horowitz:1997neutrino}) and nucleon phase-space corrections (\cite{Mezzacappa:1993gm}), reducing scattering opacities in highly degenerate media. Iso-energetic scattering on nuclear clusters $A$ (h) incorporates ion correlation, electron polarization and nuclear form factors (\cite{Burrows:2006neutrino}), while iso-energetic scattering on alpha particles is computed following (\cite{Ardevol-Pulpillo:2018btx}). Similarly, gray scattering opacities are obtained via averaging with respect to the equilibrium Fermi-Dirac distribution. As shown by (\cite{Ng:2023syk}),
and argued in (\cite{Ng:2024zve}), inelastic neutrino-electron scattering has much smaller opacity compared to iso-energetic scattering on nucleons under general BNS merger conditions. Hence, although important for the thermalization of heavy lepton neutrinos (\cite{Thompson:2000gv, Chiesa:2024lnu}), we neglect such processes, and refer to (\cite{Cheong:2024cnb}) for a single, post-merger like neutron star study.
For reactions (i) - (l) in Table~\ref{tab:reactions} (only considered for $\nu_\mu,~\bar\nu_\mu$ and $\nu_x$) we compute angle-averaged annihilation kernels, which corresponds, on one hand, to retaining the 0th Legendre moment of the kernels, and, on the other hand, to model those processes as isotropic emission/absorption. The annihilation kernel of electron-positron pair annihilation (i) is given by (\cite{Bruenn:1985, Pons:1998st}), and of nucleon-nucleon bremsstrahlung (j) from (\cite{Hannestad:1997gc}). Finally, we follow (\cite{Guo:2020tgx}) to produce isotropic annihilation kernels of inverse muon decay (k) and inverse antimuon decay (l).

Next, we describe our method for computing gray number $\kappa_{p,0}$ and energy $\kappa_{p,1}$ opacities for the pair processes (i) -(l). Most of the approaches in the literature rely on applying Kirchhoff's law to the unblocked number ($i=0$) and energy ($i=1$) emission rates $Q_{p,i}$, i.e.
\begin{equation}\label{eq:kirch}
    \kappa_{p,i} = \frac{Q_{p,i}}{c\mathcal{B}_{i}(\eta_\nu, T)},
\end{equation}
where the black-body integral reads
\begin{equation}
    \mathcal{B}_i(\eta_\nu, T) = \frac{4 \pi}{(hc)^3} T^{3 + i} F_{2 +i}(\eta_\nu),
\end{equation}
$F_{2+i}(\eta_\nu)$ is the $(2+i)$th order Fermi integral, and the neutrino degeneracy $\eta_\nu$ is set by thermal and chemical equilibrium with matter. For heavy-lepton neutrinos, $\eta_{\nu_x} = 0$ and the denominator of Eq.~\eqref{eq:kirch} never vanishes. This, however, is not the case for muon (anti)neutrinos. In particular, at low $Y_\mu$ (densities smaller than nuclear saturation), $\eta_{\nu_\mu}$ or $\eta_{\bar\nu_\mu} = -\eta_{\nu_\mu}$ often become a large negative number. In this case, the black-body integral vanishes and the associated opacity diverges. To circumvent this, \cite{Ng:2024zve} adopts a number of restrictions (see App.~\ref{sec:lit-comp}).

Instead, we propose an alternative approach, based on the construction of well-behaved pair opacities. To do so, we first note that the isotropic collisional integral for the neutrino $\nu$ undergoing a pair process is (\cite{Shibata:2011kx}
\begin{eqnarray}
    B_{\nu,p}^{(0)}= (1 - f_\nu)\int d\Omega' d\epsilon' \frac{\epsilon'^2}{(hc)^3} R^{\rm prod}_0(\epsilon, \epsilon')(1-f'_{\bar\nu}) \nonumber \\
    - f_\nu \int d\Omega' d\epsilon' \frac{\epsilon'^2}{(hc)^3}R^{\rm ann}_0(\epsilon,\epsilon') f'_{\bar\nu},\nonumber \\
    \label{eq:coll-int}
\end{eqnarray}
where ($f_{\bar\nu}$) $f_\nu$ is the (anti) neutrino distribution function, and $\Omega$ is a solid angle in spherical coordinates. For a clearer notation, we omit the functional dependencies of the distribution functions, but in general, for a spacetime coordinate $x^\mu$, $f_\nu = f_\nu(\epsilon,\Omega,x^\mu)$ and $f'_{\bar\nu} = f'_{\bar\nu}(\epsilon', \Omega', x^\mu)$, i.e., primed variables depend on primed momentum-space variables. $R^{\rm ann}_0(\epsilon,\epsilon')$ ($R^{\rm prod}_0(\epsilon,\epsilon')$) is the isotropic kernel for annihilation (production) of neutrino pairs. We note that the collisional integral has the same form for (anti)muon decay, but the pairing species are $\{\nu_{\mu},~\bar\nu_e\}$ ($\{\bar\nu_\mu, ~\nu_e\}$).

Next, as usual, we neglect neutrino blocking, and identify the spectral emissivity and opacity, respectively, as
\begin{eqnarray*}
    j_p(\epsilon) &=& \frac{4\pi}{(hc)^3}\int d\epsilon' \epsilon'^2 R^{\rm prod}_0(\epsilon, \epsilon'),\\
    \kappa_p(\epsilon) &=& \int d\Omega' d\epsilon' \frac{\epsilon'^2}{(hc)^3}R^{\rm ann}_0(\epsilon,\epsilon') f'_{\bar\nu},
\end{eqnarray*}
which allows one to re-write the collisional integral Eq.~\eqref{eq:coll-int} as
\begin{equation}
    B_{\nu, p}^{(0)} = j_p(\epsilon) - \kappa_p(\epsilon) f_\nu,
\end{equation}
and detailed-balance as
\begin{equation}\label{eq:det-bal}
    j_p(\epsilon) = \kappa_p(\epsilon) f^{\rm eq}_\nu(\epsilon, T, \eta_\nu),
\end{equation}
where $f^{\rm eq}_\nu(\epsilon, T, \eta_\nu)$ is the usual Fermi-Dirac distribution in local thermodynamical equilibrium (LTE) with matter.

Finally, the strongest approximation we make in order to produce gray pair opacities is that the pairing neutrino is in LTE, i.e., $f_{\bar\nu}' \rightarrow f'^{\rm eq}_{\bar\nu}(\epsilon', T, \eta_{\bar\nu})$~(\cite{Bollig:2017lki, Bollig:2018thesis}). We present a detailed discussion about the limitations of this approximation at the end of this Section.

In this case, the spectral pair opacity becomes
\begin{equation}\label{eq:pair-op}
    \kappa_p(\epsilon) = \frac{4\pi}{(hc)^3}\int d\epsilon' \epsilon'^2 R^{\rm ann}_0(\epsilon,\epsilon') f_{\bar\nu}'^{\rm eq}(\epsilon',T,\eta_{\bar\nu}),
\end{equation}
and we define the gray number ($i = 0$) and energy ($i=1$) opacity $\kappa_{p, i}$ for neutrino $\nu$ as
\begin{eqnarray}\label{eq:avg-op}
    \kappa_{p,i} &=& \frac{\int d\epsilon~\epsilon^{2+i} \kappa_p(\epsilon) f^{\rm eq}_\nu(\epsilon,T,\eta_\nu)}{\int d\epsilon~\epsilon^{2+i} f^{\rm eq}_\nu(\epsilon,T,\eta_\nu)} \nonumber \\
    &=& \frac{4\pi}{(hc)^3} \frac{1}{\mathcal{B}_i(T,\eta_\nu)} \int d\epsilon~\epsilon^{2+i} \kappa_p(\epsilon) f^{\rm eq}_\nu(\epsilon,T,\eta_\nu), \nonumber \\
\end{eqnarray}
which automatically satisfies the Kirchhoff's law for number and energy emission rates
\begin{equation}\label{eq:pair-emiss}
    Q_{p,i} = \int d\epsilon~\epsilon^{2+i} j_p(\epsilon) = \kappa_{p,i} ~c\mathcal{B}_i(T,\eta_\nu),
\end{equation}
when detailed-balance Eq.\eqref{eq:det-bal} is enforced. It is worth noting that this construction leads to finite pair opacities, because for very negative neutrino degeneracy
\begin{equation}
    \eta_\nu \ll -1 \rightarrow f^{\rm eq}_\nu \approx \exp(\eta_\nu) \exp(-\epsilon/T),
\end{equation}
and the $\exp(\eta_\nu)$ term factors out from the numerator and denominator of Eq.\eqref{eq:avg-op}. It also provides finite emission rates Eq.\eqref{eq:pair-emiss} for $\eta_\nu \gg 1$, because the spectral opacity $\kappa_p(\epsilon)$ is strongly suppressed by a multiplicative factor $\exp(\eta_{\bar\nu}) = \exp(-\eta_\nu) \ll 1$.

In order to avoid over (de)muonization of matter, similar to \cite{Bollig:2018thesis, Ng:2024zve}, we restrict the muonic reactions (c), (d), (f), (k) and (l) to thermodynamical conditions where muons may be found, which is implemented via a phenomenological cutoff applied to the number ($i=0$) and energy ($i=1$) absorption opacities $\kappa_{a,i}$ for those processes of the form
\begin{equation}
    \kappa_{a,i} \rightarrow \frac{\kappa_{a,i}}{[1 + (\rho_{\rm thr}/\rho)^5][1+(T_{\rm thr}/T)^6]},
\end{equation}
where $\rho_{\rm thr} = 10^{11}~{\rm g~cm^{-3}}$ and $T_{\rm thr} = 2.5~{\rm MeV}$.

Now we address whether our approach is able to capture a fundamental feature of pair-processes, namely, that the number of produced neutrinos and pairing neutrinos is equal. By our definitions Eqs.~\eqref{eq:pair-op} and \eqref{eq:pair-emiss}, the difference of number emission rates is
\begin{eqnarray}
    \Delta Q_{p,0} 
    &=& \frac{(4\pi)^2c}{(hc)^6}\iint d\epsilon~d\epsilon' \epsilon^2\epsilon'^2 R^{\rm ann}_0(\epsilon,\epsilon')\times \nonumber \\ && \hspace{2.5cm}(f'^{\rm eq}_{\bar\nu}f^{\rm eq}_\nu - f^{\rm eq}_{\bar\nu}f'^{\rm eq}_\nu),
\end{eqnarray}
which vanishes (the number emission rates are equal) if the integrand vanishes. This is automatically satisfied by heavy-lepton neutrinos, or more generally when $\eta_\nu = -\eta_{\bar\nu} =0$. Otherwise, the vanishing of the product of distribution functions is guaranteed if the isotropic kernel is symmetric on the energies. This is the case for nucleon-nucleon bremsstrahlung (j), as the kernel depends on the sum of neutrino energies. For electron-positron pair annihilation (i), the kernel is not symmetric, but \cite{Kawaguchi:2024naa} show that the symmetric part of the kernel dominates by a few orders of magnitude under relevant thermodynamical conditions. Finally, the kernel of muon and antimuon decay (reactions (k) and (l), respectively) is highly asymmetric. Thus, to ensure $\Delta Q_{p,0} =0$, we first compute the opacity for muon (anti)neutrinos using Eq.\eqref{eq:avg-op} and the corresponding emission rates $Q_{p,i}$ with Eq.\eqref{eq:pair-emiss}. Then, we equate emission rates of electronic and muonic (anti)neutrinos $\bar{Q}_{p,i} = Q_{p,i}$ and estimate the electron (anti)neutrinos opacities using Eq.~\eqref{eq:pair-emiss} with the corresponding black-body functions. The impact of this procedure is negligible for electron (anti)neutrinos, since $\beta$-reactions dominate by far the electronic gray rates.

It is worth noting that including neutrino blocking in the collisional integral and detailed-balance leads to the usual opacity correction for stimulated-absorption. In this case, emission rates still follow the integrated Kirchhoff's law, but $\Delta Q_{p,0}$ does not vanish for non-vanishing neutrino degeneracy. Hence, under the LTE approximation, inclusion of neutrino blocking leads to different number emission rates for pairing species, violating the pair nature of such processes.

The assumption that the pairing neutrino $\bar\nu$ is in LTE allows the computation of finite rates, as well as to recover the correct vanishing limit of the collisional integral should the moments of the distribution function evolve into equilibrium values. However, as shown in Eq.~(4.19) of~\cite{Shibata:2011kx}, the collisional integral for pair processes couples, in a rather complicated manner, (energy-dependent) moments of the distribution function for $\nu$ and $\bar\nu$, even in the isotropic case. Therefore, a better treatment would require the implicit solution of the moments evolution equations coupling pairs of neutrino species, which is not the goal of the present work. An alternative approach is suggested by~\cite{Fujibayashi:2017xsz}, where pair rates are evaluated using the integrated moments $J$ and $H^\mu$ of pairing species. In this form, a coupling between moments of neutrinos and anti-neutrinos in the emission rates is induced. It is unclear, however, how the moments are implicitly evolved in a self-consistent manner. We also point out that a similar LTE treatment for pair-processes for heavy-lepton neutrinos in the core-collapse supernovae simulations of \cite{OConnor:2014sgn} leads to $\sim 7\%$ differences in neutrinos luminosities and average energies compared to an energy-dependent scheme. Similarly, more recent results (\cite{Betranhandy:2025amv}) suggest an impact on the $10-15\%$ level, while differences in rest-mass density and temperature at the late post-bounce, where densities and temperatures are comparable to those of merger remnants, are very modest. It is more difficult, however, to estimate the impact for muonic neutrinos, but the uncertainty introduced by our approximation should be smaller than neglecting pair processes altogether.

We also remark that, in order to capture the trapping of a gas of neutrinos in the matter, we extend the widely adopted equilibration prescription of \cite{Perego:2019adq} to the $5$-$\nu$ case, where equilibrium timescales for electronic and muonic neutrinos are employed to determine the thermodynamical conditions at which trapping occurs. Likewise, following (\cite{Foucart:2016rxm, Foucart:2024npn}), we correct all gray opacities in the freely-streaming regime via\noindent
\begin{eqnarray}\label{eq:op-corr}
    \kappa &\rightarrow& \kappa
    ~\max\left\{1,\min\left[10, (T_\nu/T)^2\right]\right\},\\
        T_\nu &=& \frac{J}{n} \frac{F_2(\eta_\nu)}{F_3(\eta_\nu)},
\end{eqnarray}
where the neutrino temperature $T_\nu$ is estimated based on the comoving energy density $J$ and comoving number density $n$. For (partially) trapped conditions, we simply recompute gray opacities at the (partial) equilibrium thermodynamical state and apply Kirchhoff's law to obtain consistent emission rates.

Finally, the interaction rates coupling matter and radiation read\noindent
\begin{eqnarray}
    Q_i &=& Q_{\beta,i} + Q_{p,i}, \\
    \kappa_{a, i} &=& \kappa_{\beta,i} + \kappa_{p, i},
\end{eqnarray}
for energy ($i=1$) and number ($i=0$) emission rates $Q_i$ and absorption opacities $\kappa_{a,i}$, while scattering opacities $\kappa_s$ are always averaged with respect to the energy spectrum.

\section{Equilibration: the two timescales approach}\label{sec:equili}
Given the (possibly) stiff coupling between matter and radiation fields, the usage of the fluid's temperature $T$ and composition $Y_e$ (after the explicit substep) to determine interaction rates may lead to numerical instabilities, specially in very opaque regions, where radiation and matter equilibrate on timescales much shorter than an evolution timestep $\Delta t$. Hence, long-term stable BNS merger simulations with gray M1 schemes are achieved by means of equilibration prescriptions, from which the expected fluid's temperature $T^*$ and composition $Y_e^*$ after the interaction are inferred and employed to compute interaction rates~(\cite{Foucart:2024npn}). Crucially, an equilibration prescription is needed in order to properly capture the thermodynamical state of matter when a (partially) trapped gas of neutrinos is formed. Thus, in the following we review the approaches reported in the literature for the 3-$\nu$ case, and present our extension to the 5-$\nu$ case.

The equilibrium timescales for (anti)neutrinos are defined as~\cite{1986rpa..book.....R}
\begin{eqnarray}
    \tau_{\nu_l/\bar\nu_l} &=& \left[c\sqrt{\kappa_{a, \nu_l/\bar\nu_l}(\kappa_{a,\nu_l/\bar\nu_l}+\kappa_{s,\nu_l/\bar\nu_l})}\right]^{-1},
\end{eqnarray}
and quantifies how quickly (anti)neutrinos reach equilibrium via scattering and absorption on the medium. Then, the formation of a trapped gas of neutrinos and antineutrinos of flavor $l = \{e,\mu\}$ is dictated by the shortest equilibrium timescale for the lepton flavor
\begin{eqnarray}
    \tau_{l} &=& \min[\tau_{\nu_l}, \tau_{\bar\nu_l}].
\end{eqnarray}
 In previous works, only $l=e$ is considered, and the (partial) trapping of heavy-lepton neutrinos is assumed whenever electron (anti)neutrinos are (partially) trapped. Indeed, comparing the evolution timestep $\Delta t$ with the equilibrium timescale $\tau_e$, three cases arise, namely, neutrinos are considered to be: 
\begin{itemize}
    \item[(i)] Freely-streaming (F), if if $\tau_e \geq \Delta t$,
    \item[(ii)] Partially-trapped (PT), if $0.5~\Delta t <\tau_e < \Delta t$,
    \item[(iii)] Trapped (T), if $\tau_e \leq 0.5~\Delta t$.
\end{itemize}
Note that the factor $0.5 \Delta t$ is chosen to match the possibility of trapping in-between the explicit Runge-Kutta substeps.
Then, for the PT and T cases, the fluid's equilibrium temperature $T_{\rm eq}$ and equilibrium composition $Y_{e,\rm eq}$ are found by enforcing energy and lepton number conservation as solutions of the system~(\cite{Perego:2019adq})
\begin{eqnarray}
    Y_{Le} &=& Y_{e, {\rm eq}} + Y_{\nu_e}(Y_{e,\rm eq}, T_{\rm eq}) - Y_{\bar\nu_e}(Y_{e,\rm eq}, T_{\rm eq}),\label{eq:ye-eq}\\
    \varepsilon_{\rm tot} &=& \varepsilon_{\rm fl}(Y_{e,\rm eq},T_{\rm eq}) +\sum_i \varepsilon_{\nu_i}(Y_{e,\rm eq}, T_{\rm eq})\label{eq:e-eq},
\end{eqnarray}
where the fluid's rest-mass density $\rho$ is kept fixed, $\varepsilon_{\rm tot} = \varepsilon_{\rm fl}(\rho,T,Y_e) + \sum_i J_{\nu_i}$, $Y_{Le} = Y_e + (n_{\nu_e} - n_{\bar\nu_e})/n_b$, with $Y_e$ the fluid's net electron fraction, $n_{\nu_e}~(n_{\bar\nu_e})$ the evolved comoving (anti)neutrino number densitiy and $n_b$ the baryon number density. Explicit expressions for the neutrino number fractions $Y_{\nu_i}$ and energy densities $\varepsilon_{\nu_i}$ are provided in App.~\ref{app:A}. Once the equilibrium solution is found, neutrino rates are computed as functions of $(\rho, T^*, Y_e^*)$, where for state T, $Y_e^* = Y_{e,{\rm eq}},~T^* = T_{\rm eq}$, for state F, $Y_e^* = Y_e,~T^*=T$, and for state PT we linearly interpolate
\begin{eqnarray}
    Y_e^* &=& Y_{e,{\rm eq}} + (Y_e - Y_{e,{\rm eq}})\left(\frac{\tau_e}{0.5\Delta t}-1 \right),\label{eq:ye-intp} \\
    T^* &=& T_{\rm eq} + (T - T_{\rm eq})\left(\frac{\tau_e}{0.5\Delta t}-1 \right).    
\end{eqnarray}
So far, this is our approach for the 3-species equilibration scheme.

Next, we want to extend the formalism to the 5-species case, where a second equilibration timescale $\tau_\mu$ describes the state of muon (anti)neutrinos. For simplicity, we keep the assumption that heavy-lepton neutrinos become (partially) trapped with the shortest timescale, i.e., $\tau_x = \min[\tau_e,\tau_\mu]$, leading to a total of $9$ possible states (otherwise, one would have $27$ cases considering heavy-lepton neutrinos and antineutrinos independently). Naturally, the equilibrium system given by Eqs.~\eqref{eq:ye-eq}-\eqref{eq:e-eq} has to be modified in two ways: first, the functional dependencies must also include the equilibrium muon fraction $Y_{\mu,{\rm eq}}$, and second, the system has to be augmented by the muon lepton number conservation equation
\begin{equation}
    Y_{L\mu} = Y_{\mu, {\rm eq}} + Y_{\nu_\mu}(Y_{e,\rm eq}, Y_{\mu, {\rm eq}},T_{\rm eq}) - Y_{\bar\nu_\mu}(Y_{e,\rm eq}, Y_{\mu, {\rm eq}},T_{\rm eq}),\label{eq:ym-eq}
\end{equation}
with $Y_{L\mu} = Y_\mu + (n_{\nu_\mu} - n_{\bar\nu_\mu})/n_b$, $Y_\mu$ being the fluid's net muon fraction, $n_{\nu_\mu} ~(n_{\bar\nu_\mu})$ the comoving evolved (anti)neutrino number density. Again, expressions for the full equilibrium system are in App.~\ref{app:A}.  

In the following we give a detailed account of the different policies employed to determine (partially) trapped states of the fluid with two independent leptonic species. A state is hereafter referred by a duplet (X,Y), where X,Y = \{F, PT, T\}, the first entry is for electron-flavored neutrinos, and the second for muon-flavored neutrinos. As opposed to the full equilibrium system, we define a partial equilibrium system as a subsystem containing one lepton number conservation equation of the form Eq.~\eqref{eq:ym-eq} (either for $l=e$ or $l=\mu$), and a corresponding energy conservation equation of the form Eq.~\eqref{eq:e-eq}, receiving contributions only from (anti)neutrinos of flavor $l$ and heavy-lepton neutrinos. Besides, a partial equilibrium system for $l$ is to be solved at fixed number fraction of the other flavor. For example, a partial system for $l=e$ is to be solved for $T_{\rm eq},~Y_{e,{\rm eq}}$ at fixed $Y_\mu$, while a partial system for $l=\mu$ is to be solved for $T_{\rm eq},~Y_{\mu,{\rm eq}}$ at fixed $Y_e$.

Finally, we reserve the notation $I_l(Z_{\rm eq}, Z)$ for a linear interpolation of the form Eq.~\eqref{eq:ye-intp} with $\tau_l$ for any fluid variable $Z$ with equilibrium value $Z_{\rm eq}$.

The simplest possible case is (F, F), where no equilibrium system has to be solved, thus $T^* = T,~Y_e^* = Y_e,~Y_\mu^* = Y_\mu$. Next, for state (T, T) we solve the the full equilibrium system and set $T^* = T_{\rm eq},~Y_e^* = Y_{e,{\rm eq}}, ~Y_\mu^* = Y_{\mu,{\rm eq}}$. For (T, F)/(F, T), we solve the partial system for $l=e/\mu$ and set $T^* = T_{\rm eq},~Y_e^* = Y_{e,{\rm eq}}/Y_e,~Y^*_\mu = Y_\mu/Y_{\mu,{\rm eq}}$, with the rationale that a freely-streaming species will interact with matter that equilibrates with the other species, and that the lepton number fractions are independent. Likewise, for (PT, F)/(F, PT), we solve the partial system for $l=e/\mu$ and make $T^* = I_{e/\mu}(T_{\rm eq},T),~Y_e^* = {\rm I}_e(Y_{e,{\rm eq}},Y_e)/Y_e,~Y^*_\mu = Y_\mu/I_\mu(Y_{\mu,{\rm eq}},Y_\mu)$.

For (PT, PT) we solve the full system, make $Y^*_e = I_e(Y_{e,{\rm eq}},Y_e)$, $Y^*_\mu = I_\mu(Y_{\mu,{\rm eq}},Y_\mu)$ and solve Eq.~\eqref{eq:e-eq} for $T^*(Y_e^*, Y_\mu^*)$ at fixed $Y_e^*,~Y_\mu^*$. Finally, for (T, PT)/(PT, T), we solve the full system, set $Y^*_e = Y_{e,{\rm eq}}/I_e(Y_{e,{\rm eq}},Y_e)$, $Y^*_\mu = I_\mu(Y_{\mu,{\rm eq}},Y_\mu)/Y_{\mu, {\rm eq}}$ and, similar to the previous case, solve Eq.~\eqref{eq:e-eq} for $T^*(Y_e^*, Y_\mu^*)$ at fixed $Y_e^*,~Y_\mu^*$. For convenience, we summarize our scheme in Table~\ref{tab:eq-scheme}.

\begin{table}[t!]
    \centering
        \caption{Summary of the two timescales equilibration scheme. The columns read, from left to right, state of the neutrino pairs, the first entry corresponding to $\nu_e,~\bar\nu_e$ and the second to $\nu_\mu,~\bar\nu_\mu$ (see text for the definitions of T, PT and F); equilibrium system, fluid electron fraction, fluid muon fraction and fluid temperature after interaction.}
    \begin{tabular}{c|c|c|c|c}
        State &  Eq. System & $Y_e^*$ & $Y_\mu^*$ & $T^*$\\
        \hline \hline
        (F, F) &  -- & $Y_e$ & $Y_\mu$ & $T$\\
        (T, T) & Full & $Y_{e, {\rm eq}}$ & $Y_{\mu, {\rm eq}}$ & $T_{\rm eq}$\\
        (T, PT) & Full & $Y_{e, {\rm eq}}$ & $I_\mu(Y_{\mu,{\rm eq}},Y_\mu)$ & $T^*(Y_e^*, Y_\mu^*)$\\
        (PT, PT) & Full & $I_e(Y_{e,{\rm eq}},Y_e)$ & $I_\mu(Y_{\mu,{\rm eq}},Y_\mu)$ & $T^*(Y_e^*, Y_\mu^*)$\\
        (PT, T) & Full & $I_e(Y_{e,{\rm eq}},Y_e)$ & $Y_{\mu, {\rm eq}}$ & $T^*(Y_e^*,Y_\mu^*)$\\
        (T, F) & Partial $l=e$ & $Y_{e, {\rm eq}}$ & $Y_\mu$ & $T_{\rm eq}$\\
        (PT, F) & Partial $l=e$ & $I_e(Y_{e,{\rm eq}},Y_e)$ & $Y_\mu$ & $I_e(T_{\rm eq}, T)$\\
        (F, PT) & Partial $l=\mu$ & $Y_e$ & $I_\mu(Y_{\mu,{\rm eq}},Y_\mu)$ & $I_\mu(T_{\rm eq},T)$\\
        (F,T) & Partial $l=\mu$ & $Y_e$ & $Y_{\mu, {\rm eq}}$ & $T_{\rm eq}$
    \end{tabular}%
    \label{tab:eq-scheme}
\end{table}

Once $(\rho, T^*, Y_e^*,Y_\mu^*)$ are obtained, we recompute opacities in the PT and T states accordingly, while neutrinos in the F state have opacities corrected as in Eq.~\eqref{eq:op-corr}.
Note that this correction follows the prescription of \cite{Foucart:2016rxm}, and is based on the approximate $\epsilon^2$ scaling of opacities in the elastic approach of, e.g., \cite{Ruffert:1995fs}. In optically thin medium (i.e., at low densities), where the various corrections from full kinematics are negligible, the elastic limit is recovered, and such a correction is reasonable. However, at intermediate densities and temperatures, where neutrinos are typically in the PT state, the scaling with energy significantly deviates from a quadratic, and the usual interpolation based on equilibrium timescales would be inaccurate.

Finally, from the corrected opacities, we obtain emission rates from Kirchhoff's law, therefore constructed to capture the (partial) trapping of radiation and matter with the correct equilibrium state of matter.
\section{Equilibrium Systems}\label{app:A}
In this section we provide explicit expressions for the equilibrium equations used in Sec.~\ref{sec:equili}. For the 3-species case, they read (at fixed baryon number density $n_b$)~(\cite{Perego:2019adq})
\begin{eqnarray}
    Y_{Le} &=& Y_{e, {\rm eq}} + \frac{4\pi}{3(hc)^3n_b} T_{\rm eq}^3\eta_{\nu_e}(\pi^2 + \eta_{\nu_e}^2), \label{eq:eq-ye}\\
    \varepsilon_{\rm tot} &=& \varepsilon_{\rm fl}(Y_{e,{\rm eq}},T_{\rm eq}) \nonumber \\
    &+& \frac{4\pi}{(hc)^3} T_{\rm eq}^4\left[\frac{21\pi^4}{60} +\frac{1}{2}\eta_{\nu_e}^2\left(\pi^2 + \frac{1}{2}\eta_{\nu_e}^2\right)\right]\label{eq:eq-e},
\end{eqnarray}
where $\eta_{\nu_e} = \eta_{\nu_e}(Y_{e,{\rm eq}},T_{\rm eq})$ is the electron neutrino degeneracy. For those expressions, we used~(\cite{1978ApJ...224..631B})
\begin{eqnarray*}
    Y_{\nu_e} - Y_{\bar\nu_e} &=& \frac{4\pi}{(hc)^3n_b} T^3(F_2(\eta_{\nu_e}) - F_2({-\eta_{\nu_e}})), \\
    \varepsilon_{\nu_e} + \varepsilon_{\bar\nu_e} + \varepsilon_{\nu_x} &=& \frac{4\pi}{(hc)^3} T^4(F_3(\eta_{\nu_e}) + F_3({-\eta_{\nu_e}})
   +4F_3(0)), \\
    F_2(\eta_{\nu_e}) - F_2(-\eta_{\nu_e}) &=& \frac{1}{3}\eta_{\nu_e}(\pi^2 + \eta_{\nu_e}^2),\\
     F_3(\eta_{\nu_e}) +F_3(-\eta_{\nu_e}) &=& \frac{7\pi^4}{60} +\frac{1}{2}\eta^2_{\nu_e}\left(\pi^2 + \frac{1}{2}\eta_{\nu_e}^2\right),\\
     F_3(0) &=& \frac{7\pi^4}{120}.
\end{eqnarray*}
Now, for the 5-$\nu$ case, the expressions for the partial equilibrium systems are obtained by substituting the numerical factors $21 \rightarrow 14$, reflecting the fact that heavy-lepton neutrinos represent 2 species, instead of 4. With this adaption, the partial $l=e$ equilibrium system is of the form Eqs.~\eqref{eq:eq-ye}-\eqref{eq:eq-e}, to be evaluated at fixed $Y_\mu$. Likewise, the partial $l=\mu$ system is obtained by changing $Y_e \rightarrow Y_\mu$, $\eta_{\nu_e} \rightarrow \eta_{\nu_\mu}$, to be evaluated at fixed $Y_e$. It is also worth noting that in the cases of partial equilibrium, the temperature $T^*(Y_e^*, Y_\mu^*)$ is obtained by solving Eq.~\eqref{eq:eq-e} for $T_{\rm eq}$ at fixed $Y_e^*, Y_\mu^*$ appropriately set by the radiation state.

Finally, the full equilibrium system reads
\begin{eqnarray}
    Y_{Le} &=& Y_{e, {\rm eq}} + \frac{4\pi}{3(hc)^3n_b} T_{\rm eq}^3\eta_{\nu_e}(\pi^2 + \eta_{\nu_e}^2), \label{eq:eq-ye-5}\\
       Y_{L\mu} &=& Y_{\mu, {\rm eq}} + \frac{4\pi}{3(hc)^3n_b} T_{\rm eq}^3\eta_{\nu_\mu}(\pi^2 + \eta_{\nu_\mu}^2), \label{eq:eq-ym-5}\\ 
    \varepsilon_{\rm tot} &=& \varepsilon_{\rm fl}(Y_{e,{\rm eq}}, Y_{\mu,{\rm eq}}, T_{\rm eq}) \nonumber \\
    &+& \frac{4\pi}{(hc)^3} T_{\rm eq}^4\left[\frac{21\pi^4}{120} +\frac{1}{2}\eta_{\nu_e}^2\left(\pi^2 + \frac{1}{2}\eta_{\nu_e}^2\right)\right] \nonumber\\ 
     &+& \frac{4\pi}{(hc)^3} T_{\rm eq}^4\left[\frac{21\pi^4}{120} +\frac{1}{2}\eta_{\nu_\mu}^2\left(\pi^2 + \frac{1}{2}\eta_{\nu_\mu}^2\right)\right],\label{eq:eq-e-5}
\end{eqnarray}
where $\eta_{\nu_e} = \eta_{\nu_e}(Y_{e,{\rm eq}}, Y_{\mu, {\rm eq}}, T_{\rm eq})$, $\eta_{\nu_\mu} = \eta_{\nu_\mu}(Y_{e,{\rm eq}}, Y_{\mu, {\rm eq}}, T_{\rm eq})$, and, for the sake of homogeneity, we absorbed half of the $\nu_x$ contribution in the second and third lines of Eq.~\eqref{eq:eq-e-5}.

\section{Literature Comparison}\label{sec:lit-comp}
In this Section, we comment on the differences between our results and those of \cite{Ng:2024zve}. There, a key conclusion is that, due to muonic reactions in the remnant, matter gets strongly deprived of energy that would be otherwise used for emission of electron (anti)neutrinos. Hence, with the suppression of electronic reactions, due to cooler remnants, substantial fractions of very neutron-rich ejecta, and the absence of outflows with $Y_e \gtrsim 0.3$, are observed. To support this scenario, their analysis is based on extracting neutrino chemical potentials from the simulations, which is equivalent to our procedure of verifying equilibration of matter and radiation with an auxiliary EOS for an equilibrium mixture. Indeed, following the approach of \cite{Espino:2023dei}, we obtain good quantitative agreement with respect to the chemical potentials reported in \cite{Ng:2024zve}, except that in our 5-$\nu$ runs, chemical potentials much closer to zero are reached in the polar region.

On the other hand, important differences regarding the determination of neutrino interaction rates are to be noted: in order to avoid divergent opacities for muonic neutrino pair-processes, the authors rescale neutrino degeneracies according to the energy optical depth $\tau_\nu$, a solution of the Eikonal equation $|\nabla \tau_\nu | = \kappa_{a, \nu}$, where $\kappa_{a,\nu}$ is the energy-averaged absorption opacity for the neutrino $\nu$. Provided a boundary condition, e.g., $\tau_\nu = 0$ at the numerical domain boundary, an iterative solution may be obtained. Furthermore, in order to single out a physical solution at point $\vec{x}$, the iterative method usually incorporates Fermat's principle
\begin{equation}
    \tau_\nu(\vec{x})= \min \left[\int_\Gamma \kappa_{a,\nu}(\vec{x}')~ds\right],
\end{equation}
where the minimum line integral is taken over all paths $\Gamma$ connecting the point $\vec{x}$ to the boundary using the line element $ds$. This requirement ensures that the optical depth represents the minimum number of interactions a neutrino would experience when escaping from its current position. Intuitively this means that the deeper within a matter distribution a point is, the higher should be its optical depth. Equivalently, for the simplified geometry of an axisymmetric remnant, the optical depth should be a monotonically decreasing function of the distance to the center. This is shown, for instance, in \cite{Palenzuela:2022kqk} for rotating, hot NSs, resembling the state of a remnant after a few dynamical timescales.

The degeneracy rescaling adopted by \cite{Ng:2024zve} reads $\eta_\nu = \eta_\nu^{\rm eq}(1-e^{-\tau_\nu})$, where $\eta_\nu^{\rm eq}$ is determined assuming LTE at the fluid's thermodynamic state $(n_b, T,Y_e,Y_\mu)$ (or without $Y_\mu$ for $3$-$\nu$). For muon (anti)neutrinos only, the condition is augmented by $\eta_{\nu_\mu} = 0$ ($\eta_{\bar\nu_\mu} = 0$) for $\tau_{\nu_\mu} < 1$ ($\tau_{\bar\nu_\mu} < 1$), and $\eta_{\nu_\mu} = \eta_{\bar\nu_\mu} = 0$ for $Y_\mu < 5\times10^{-3}$. Whereas, in the present work, this is not needed, as the opacities are finite by construction.

Thus, in their approach, the neutrino degeneracy approaches $\eta_\nu \rightarrow 0$ outside of the neutrinosphere $\tau_\nu < 1$, while corrections become negligible inside, where $\tau_\nu \geq 1$. But, since opacities for gray schemes are tabulated assuming LTE, this means that $\kappa(n_b,T,Y_e,Y_\mu)$ computed from the fluid state is averaged with a FD distribution with $\eta_\nu^{\rm eq}$ that does not necessarily match $\eta_\nu$, which possibly introduces an inconsistency. Then, emissivities would be obtained by using Kirchhoff's law with $\eta_\nu$, which is potentially problematic, as we elucidate in the following.

For electron (anti)neutrinos, the $\bar\nu_e$ neutrinosphere is generally located deeper in the remnant than of $\nu_e$ (\cite{Endrizzi:2019trv}). This is specially relevant for electronization of matter in, e.g., the disk, because it generally amounts to a larger emission of $\bar\nu_e$ compared to $\nu_e$. Moreover, in those intermediate density regions, electron (anti)neutrino degeneracies assume values $\eta_{\bar\nu_e}^{\rm eq} = -\eta^{\rm eq}_{\nu_e} = [0,2]$ (\cite{Loffredo:2022prq}). Thus, given the degeneracy asymmetry between $\bar\nu_e$ and $\nu_e$, excess emission of $\bar\nu_e$ compared to $\nu_e$ may be explained, as the number emission rate of the former, proportional to the second-order Fermi integral $F_2(\eta_{\bar\nu_e})$, may become larger than the number emission rate of the latter, proportional to $F_2(\eta_{\nu_e})$, even in the typical case where the averaged opacities obey $\kappa_{a,\bar\nu_e} < \kappa_{a,\nu_e}$. This phenomenology is consistent with early electronization of polar material, reported in a myriad of works, both with leakage schemes (\cite{Sekiguchi:2010fh, Foucart:2015gaa, Lehner:2016lxy, Bovard:2017mvn, Foucart:2015vpa, Radice:2016dwd, Radice:2018pdn, Palenzuela:2022kqk}) and M1 schemes (\cite{Foucart:2016rxm, Radice:2021jtw, Schianchi:2023uky, Musolino:2023pao, Gieg:2025ivb, Neuweiler:2025klw, Daszuta:2026szb}) (see also \cite{Zappa:2022rpd} for a cross-comparison study of different neutrino transport prescriptions).

To illustrate our point, we consider typical values $\eta_{\bar\nu_e}^{\rm eq} = -\eta^{\rm eq}_{\nu_e} = \{1,~2\}$ encountered in the polar region. In this case we have, respectively, $F_2(\eta^{\rm eq}_{\bar\nu_e}) = \{4.3,~9.5\}$ and $F_2(\eta^{\rm eq}_{\nu_e}) = \{0.71,~0.27\}$, while $F_2(0) \approx 1.8$. Now, due to the aforementioned rescaling of degeneracies, $\eta_{\bar\nu_e}$ ($\eta_{\nu_e}$) would be rescaled towards values much closer to zero than to $\eta^{\rm eq}_{\bar\nu_e}$ ($\eta^{\rm eq}_{\nu_e}$), since in the polar region both neutrinospheres coincide and intercept the remnant at $\rho \approx 10^{11} - 10^{12}~{\rm g~cm^{-3}}$. Thus, electron antineutrino number emissivities may be underestimated by factors of $F_2(\eta^{\rm eq}_{\bar\nu_e})/F_2(0)$ in the range $2.4 - 5.3$, while the ratios $F_2(\eta^{\rm eq}_{\bar\nu_e})/F_2(\eta^{\rm eq}_{\nu_e})$, expressing the asymmetry of number emissivities, would be reduced from $6 - 34$ to $2.5 - 6.7$.

In this scenario, early electronization of polar material is less notable, as evident in their Figures 3 and 10, also for the $3$-$\nu$ case. For their reported $5$-$\nu$ simulations, the effect is stronger, as the neutrinospheres are expected to be located even deeper in cooler remnants. Indeed, the high $\bar\nu_e$ chemical potential reported in the upper, right panel of their Figure 2 corroborates our interpretation that electronization via excess emission of $\bar\nu_e$ seems absent because of the degeneracy rescaling. Then late time electronization is reported for their $3$-$\nu$ runs, consistent with the arise of neutrino winds, but noticeably weaker in the $5$-$\nu$ case. It is important to highlight, however, that our argument does not apply to the disk regions, where a consistent description of electronization is further complicated by non-local effects, e.g., advection of moderate $Y_e$ material from within the neutrinospheres, where rescaling is absent, and absorption of neutrinos coming from the hot remnant. Consequently, $Y_e \sim 0.25 - 0.35$ is reported for their SFHo $3$-$\nu$ disk, higher than our $Y_e \sim 0.20 - 0.25$, which we attribute to their higher maximum temperatures $T_{\rm max} \gtrsim 40~{\rm MeV}$, compared to our $T_{\rm max} \leq 40~{\rm MeV}$ already at $t-t_{\rm mrg} \approx 7~{\rm ms}$. Such differences are plausibly explained by the different pair processes treatment.

A similar effect of the degeneracy rescaling can be seen in their reported $Y_\mu$, where values in excess of $5\times10^{-3}$ are present in regions with $\rho \leq 10^{11}~{\rm g~cm^{-3}}$ and temperatures $T \lesssim 10~{\rm MeV}$, where vanishing muon fractions would be expected on a neutrinoless state driven to equilibrium predominantly by free emission of $\nu_\mu$. This is a consequence of large, negative degeneracies for muons/antimuons at low fractions. Here we state that, whether such degeneracies are physically reasonable (see~\cite{Bollig:2018thesis}) and strategies to remedy this remain open questions.

Finally, it is also worth noting that \cite{Foucart:2015vpa} argues that the degeneracy rescaling at low optical depths is of little importance for merger simulations, but possibly significant in other contexts, such as core-collapse supernovae (see also \cite{Cheong:2024buu}). It might be computationally feasible to retain consistency between averaged neutrino interaction rates and rescaled degeneracies in a few cases, e.g., if the elastic rates of \cite{Ruffert:1995fs} are employed, since the difficulty in computing opacities and emissivities is the evaluation of Fermi integrals, which can be efficiently done via accurate rational approximants (see~\cite{1978A&A....67..185T}). However, for full kinematics, tabulated gray interaction rates, it seems prohibitive for the current infrastructure of \texttt{BAM} to average spectral rates on-the-fly with rescaled degeneracies (see \cite{Chiesa:2024lnu} for an approach along those lines), and, what is more, to do so iteratively along with the computation of optical depths.

Instead, we opt for the simplest way to preserve consistency between tabulated interaction rates under LTE and the dynamical evolution of neutrino fields, which is present in our proposed method by construction. Further quantification of the uncertainties introduced by LTE treatment of rates and degeneracies is a topic for future works.

\end{document}